\documentclass[a4paper,fleqn,useAMS,usenatbib]{mnras}

\pdfoutput=1

\usepackage{natbib}
\usepackage{mathptmx}

\usepackage[T1]{fontenc}
\usepackage{ae,aecompl}

\usepackage[dvips]{graphicx}
\usepackage{amsmath}
\usepackage{amsfonts}
\usepackage{amssymb}
\usepackage{comment}
\usepackage{bm} 
\usepackage[dvipsnames]{xcolor}
\usepackage[]{hyperref}
\hypersetup{
	colorlinks=true,	
	urlcolor=MidnightBlue,		
	hypertexnames=true,
	plainpages=false,
	naturalnames=true,
	pdftitle={Astrometric identification of nearby binary stars II},    
	pdfauthor={Zephyr Penoyre},     
	linkcolor=WildStrawberry,          
	citecolor=Periwinkle,        
}

\usepackage{amsmath}
\usepackage{amsfonts}
\usepackage{amssymb}
\usepackage{wasysym}
\usepackage[percent]{overpic}
\usepackage{subfig}
\usepackage{csvsimple}
\usepackage{lscape}
\usepackage[shortlabels]{enumitem}

\makeatletter
\def\env@matrix{\hskip -\arraycolsep 
  \let\@ifnextchar\new@ifnextchar
  \array{*{\c@MaxMatrixCols}c}}
\makeatother

\title[Astrometric identiﬁcation of nearby binary stars II]{Astrometric identification of nearby binary stars II: Astrometric binaries in the Gaia Catalogue of Nearby Stars}
\author[Z. Penoyre et al.]{Zephyr Penoyre$^{1}$\thanks{E-mail:
\href{mailto:zpenoyre@ast.cam.ac.uk}{zpenoyre@ast.cam.ac.uk}}, Vasily Belokurov$^{1}$, N. Wyn Evans$^{1}$ \\
$^{1}$Institute of Astronomy, University of Cambridge, Madingley Road, Cambridge, CB3 0HA, United Kingdom}
\date{Accepted . Received ; in original form 9$^{th}$ August 2021}

\pubyear{2022}

\begin{document}
\label{firstpage}
\pagerange{\pageref{firstpage}--\pageref{lastpage}}
\maketitle

\begin{abstract}
We examine the capacity to identify binary systems from astrometric deviations alone. We apply our analysis to the \textit{Gaia} eDR3 and DR2 data, specifically the \textit{Gaia Catalogue of Nearby Stars}. We show we must renormalize (R)UWE over the local volume to avoid biasing local observations, giving a Local Unit Weight Error (LUWE). We use the simple criterion of LUWE>2, along with a handful of quality cuts to remove likely contaminants, to identify unresolved binary candidates. We identify 22,699 binary candidates within 100 pc of the Sun (just under 10\% of sources in this volume). We find an astrometric binary candidate fraction of around 20\% for giant stars, 10\% on the Main Sequence and lower than 1\% for White Dwarfs. We also look for Variability Induced Movers, by computing the correlation between photometric variability and astrometric noise - and show that VIMs may dominate the binary population of sub-Solar mass MS stars. We discuss the possibility and limitations of identifying non-luminous massive companions from astrometry alone, but find that our method is insensitive to these. Finally, we compare the astrometric deviations of MS binaries to the simulated sample from paper I, which show excellent agreement, and compute the astrometric candidate binary fraction as a function of absolute magnitude.
\end{abstract}


\section{Introduction}

Precise astrometric information allows us to build accurate models of the motion of astronomical bodies, constraining distant stars as well as our own planet and Solar System. Imprecise astrometric information can be almost as informative, specifically if we believe the source of noise to be a physical effect. 

In \citet{Penoyre20} and \citet{Belokurov20} (B+20) we have shown that the precision of the \textit{Gaia surveys'} astrometric measurement is sufficient to resolve the motion of many binary systems. Though we do not have the full astrometric time series data yet we can use the single-body astrometric solution\footnote{where each star is described by a 2D position and proper motion, and its parallax ($\varpi$)} and specifically its error to separate single stars and binary candidates. The extra motion caused by a binary with a period comparable to that of the survey can bias the apparent position and motion of a star, and introduce a significant extra noise contribution. 

The \textit{Gaia} survey \citep{Gaia16,Gaia18,Gaia21} provides astrometric measurements at very high precision for just under 2 billion stars. We focus specifically on the reduced-chi-squared value from the single-body fit, denoted in the \textit{Gaia} literature as a unit weight error ($UWE$). Values close to unity suggest a reasonable agreement between the model and observations, whilst significantly larger values will be seen if there is an extra unmodelled component of the motion, most ubiquitously a binary or higher multiple.

In paper I of this two-part series (\citealt{Penoyre21}, P+21 hereafter) we showed that main sequence (MS) binary systems within 100 pc can be expected to be detected reliably if the period is between $\sim$1 month and $\sim$ 30 years. We can use the simple but powerful estimation of the distribution of known binary periods from \citet{Raghavan10} ($\log(P)=2.47\pm2.28$ where $P$ is the period in years) to estimate the expected fraction of all binaries which may be detected astrometrically. 27\% of binaries fall within the above period range, suggesting that if $\sim$20-30\% of observed systems show evidence of being astrometric binaries then this is consistent with approximately all of them being binaries. Whilst we don't expect this to be true in general it is feasible for some populations, specifically more massive stars \citep{Duchene13,Offner20} and some regions of parameter space only attainable by the blending of two sources.

The aim of this series is to provide simple and robust criteria for selecting likely astrometric binary candidates. Thus, in this paper, we aim to set out and justify minimal criteria to separate likely binary systems from both single stars and spurious astrometric solutions. In Section \ref{gcnsbinaries}, we introduce the \textit{Gaia Catalogue of Nearby Stars} (GCNS) \citep{Smart20} and show how and why we renormalise the UWE for this local sample. In section \ref{astrometricdeviations} we show the astrometric behaviour of all GCNS stars which pass our quality cuts. Then in section \ref{binarycandidates} we identify a large number of candidate binaries, and we try to quantify the degree of contamination of the sample. We also examine the degree to which astrometric noise may be caused by variability of a longer period binary system, and explore the possibility of detecting dark companions in this local sample.

\section{Data}
\label{gcnsbinaries}

The Gaia catalogue of Nearby Stars (GCNS, \citealt{Smart20}) contains over 300,000 stars which reside within 100 pc from our solar system, selected from the full Gaia eDR3 sample \citep{Lindegren21}. The selection criteria uses a random forest classifier, which assigns a likelihood to each source of it being a valid target within 100 pc based on the full set of astrometric parameters reported in eDR3. The random forest is trained to identify sources to be removed from the sample using a set of sources with clearly spurious astrometric solutions ($\varpi<-8$ mas). 

The major contaminant in a distance limited sample is a significant number of distant stars with a spurious parallax, due to a poorly converged or otherwise pathological astrometric fit -- and this classifier is effective at removing these sources. Good evidence for this is provided by the fact that the sources deemed to be good are almost isotropic on-sky (as we'd expect for such a nearby sample) whilst the sources deemed spurious clearly map out the Galactic disk and even the Large and Small Magellanic Clouds (i.e. the distant Galaxy).

This sample is an excellent test set for exploring binary classification because it is
\begin{itemize}
\item clean - having most contaminating and nonsensical sources removed already by the random forest leaves a catalogue for which any clear trends or behaviours can be assumed to be astrophysical
\item close - these are the nearest sources to our Sun, and includes some of the most observed and best classified objects in the sky. They are of a relatively uniform and well known composition, many of which share a broadly similar history to that of our Sun
\item concise - the sheer number of stars in the full Gaia catalogue, whilst both scientifically useful and a huge technological achievement, can be overwhelming and unwieldy to work with. A much reduced catalogue such as this still covers a broad and representative range of sources
\end{itemize}

\subsection{Preparing the data}

\begin{figure}
\centering
\includegraphics[width=0.49\textwidth]{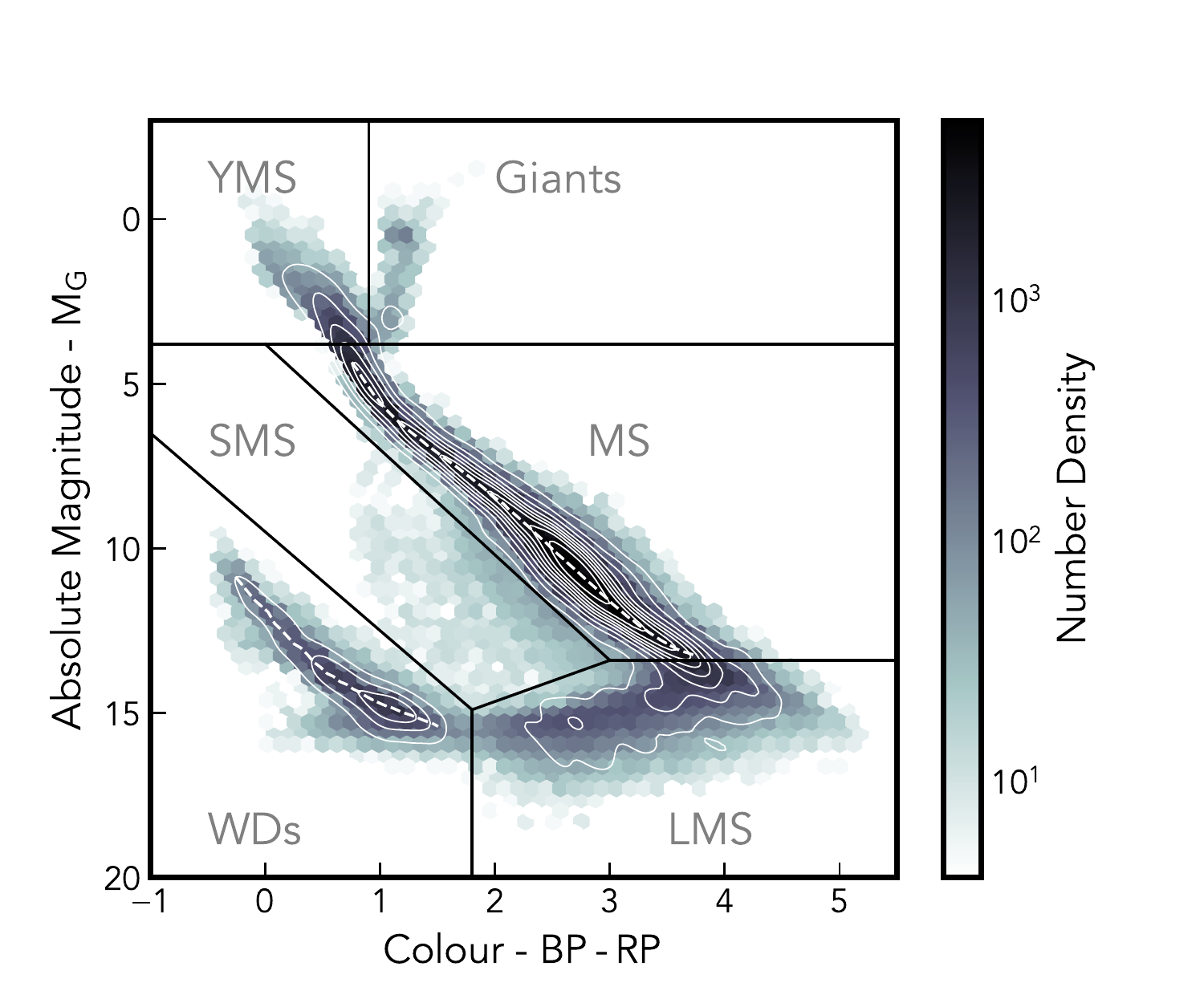}
\caption{Hertzsprung-Russel (HR) diagram showing the distribution of all sources in the GCNS. We have separated the space into 6 regions corresponding approximately to different classes of object: Lower Main Sequence (LMS), Main Sequence (MS), Young Main Sequence (YMS), Giants, White Dwarfs (WDs) and the Sub-Main Sequence (SMS). We also show a kernel density estimate of the population which meets our quality cuts in thin white lines (each contour level containing 10\% of sources). The median absolute magnitude, as a fraction of colour, is shown for MS stars and WDs as a dotted white line. Bins with less than 4 sources are considered too sparse to be meaningfully measured and are not coloured, as will be true throughout this paper.}
\label{hrSimple}
\end{figure}

In Figure \ref{hrSimple}, we show the Hertzsprung-Russel (HR) diagram for all stars in GCNS. We compute the absolute magnitude via
\begin{equation}
M_G = m_G + 5 \log_{10}(\varpi) -10
\end{equation}
where $m_G$ is the observed apparent magnitude and $\varpi$ is the parallax\footnote{A more thorough approach might account for the fact that the modal parallax does not correspond to the modal distance ($d=\frac{1}{\varpi}$) as discussed in for example \citet{BJ21}.} (in milli-arcseconds - $mas$)).
We split the space into 6 classes of source based on the observed colour ($BP-RP$) and absolute magnitude ($M_G$):
\begin{itemize}
\item Main Sequence (MS) - the most populated region filled with stars up to $\sim$the mass of our Sun which have not yet exhausted their supply of Hydrogen, with $3.8<M_G<13.4$ and $M_G-3.2(BP-RP)<3.8$
\item Young Main Sequence (YMS) - a mixture of massive young stars still able to fuse Hydrogen and various evolved but hot (and thus blue) sub-populations, with $M_G<3.8$ and $(BP-RP)<0.9$
\item Giants - evolved stars which have greatly expanded outer envelopes, becoming both much brighter and significantly cooler (redder), with $M_G<3.8$ and $(BP-RP)>0.9$
\item White Dwarfs (WDs) - compact remnants of evolved stars whose outer envelope has been completely stripped leaving just a small hot degenerate core, with $M_G-3(BP-RP)>9.5$ and $(BP-RP)<1.8$
\item Lower-Main Sequence (LMS) - a mixture of low mass MS stars and lower mass objects which may be unable to fuse hydrogen even at their core, whose luminosity mostly comes from gravitational contraction, with $(BP-RP)>1.8$, $M_G>13.4$ and $ M_G+\frac{4}{3}(BP-RP)>17.4$
\item Sub-Main Sequence (SMS) - a sparsely populated region between the MS and WDs, consisting of a mixture of MS+WD binaries and (as we will argue in this section) a high fraction of spurious sources with incorrect parallax measurements. Occupying the region between LMS, MS and YMS stars and WDs.
\end{itemize}

These boundaries have been chosen with reference to the HR diagram of GCNS, and are intended to be very simplistic. Many sub-populations, especially of evolved/bright stars are grouped together, and some (such as the Horizontal branch) even split between categories. 

The LMS group (and dim red sources in general) also suffers from low signal levels with measurements of $BP$ (and for studies of this region it is suggested that $(G-RP)$ is a more representative colour metric \citealt{Riello21,Fabricius21}) hence we will mostly exclude them from further analysis. This is the reason they appear to be bending towards the WDs and not a clean continuation of the MS.

For the MS and WD regions, we also calculate a median absolute magnitude as a function of colour, for the population, $ \langle M_G \rangle(BP-RP)$. This allows us to find the offset between any given MS or WD source from their respective main sequences:
\begin{equation}
\label{damag}
\delta M_G = M_G-\langle M_G \rangle(BP-RP)
\end{equation}
which will be negative if sources are brighter than the median.

The sample of stars we are examining here is relatively complete, due to the small distance cut of 100 pc. Thus, the giant branch is much less apparent than samples of more distant regions (such as globular clusters) or samples without any distance cut - where in both cases the low mass stars are suppressed by their low apparent magnitude.

For each source in the data, we find the associated source in the DR2 catalogue, and apply 6 quality cuts to remove sources which are poorly resolved, lacking data or otherwise suspect (the largest contributor being sources in crowded regions). We explore this process in detail in Appendix \ref{qualityCuts}. These cuts remove just over 20\% of the sample, leaving us with 260,164 sources.

\begin{figure*}
\centering
\includegraphics[width=0.98\textwidth]{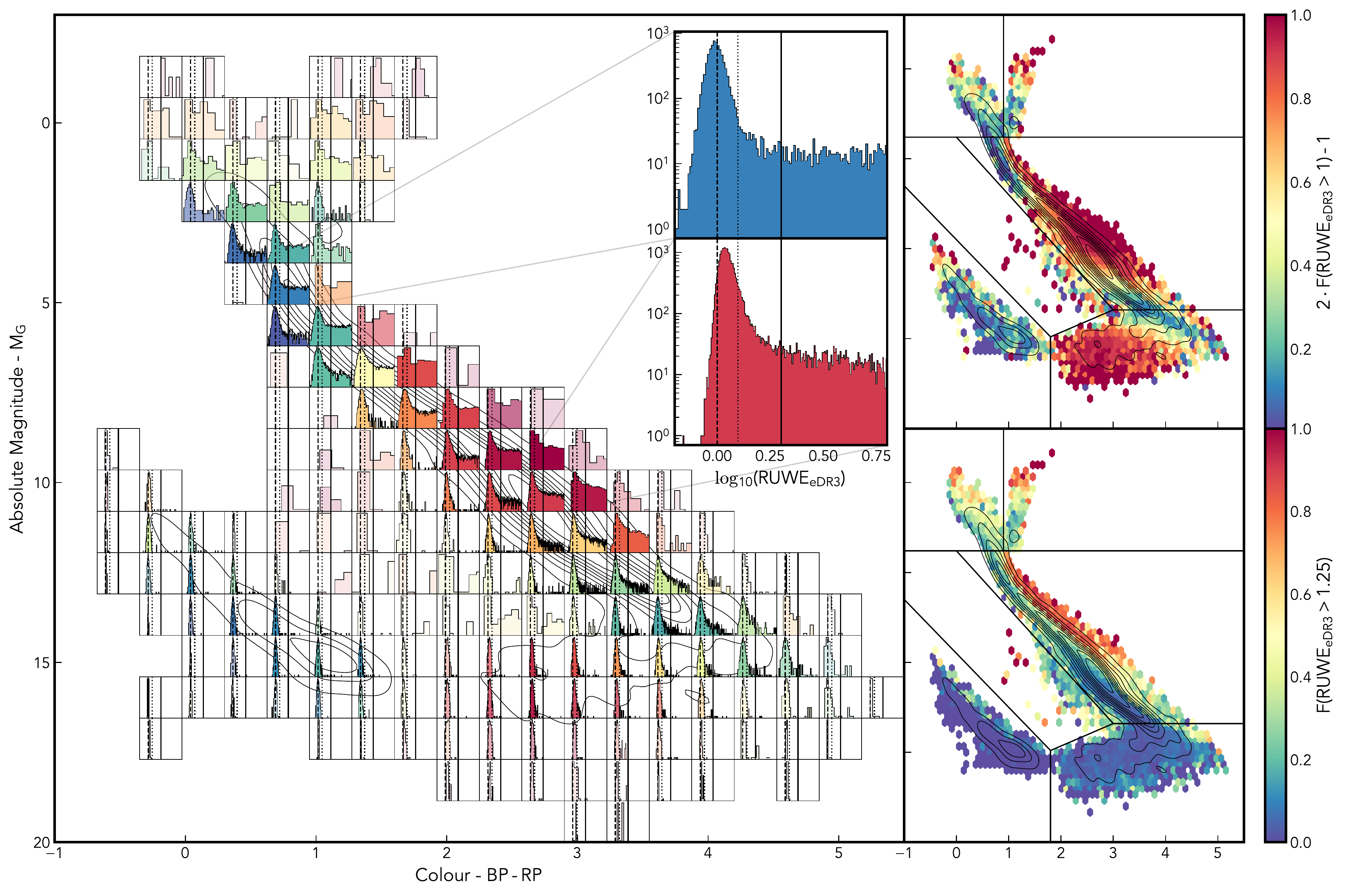}
\caption{Left - the distribution of $RUWE$ is shown for populations based on their position on the HR diagram. Each histogram shows the RUWE distribution for every source contained in that box. Boxes with 3 or less sources are excluded, and the opacity of histograms increases with number of sources contained (up to 1000). $RUWE$s of 1, 1.25 and 2 are shown as dashed, dotted and solid vertical lines. The colour of each plot is set by the fraction of sources with $RUWE>1$ (using the same colouring as the top right panel), a measure of the degree of systematic overestimation of $RUWE$. The inset panels show the distribution for Sun-like stars (upper) and M-type stars (lower). Top right - The fraction of sources with $RUWE>1$, normalised such that a well behaved population of single stars should have a value of 0, and populations dominated by sources with high $UWE$ tend to 1. Bottom right - The fraction of sources with $RUWE>1.25$. The boundaries between classifications of source and the density of sources throughout the whole sample are repeated from Figure \ref{hrSimple}.}
\label{hrRuweHists}
\end{figure*}

\begin{figure*}
\centering
\includegraphics[width=0.98\textwidth]{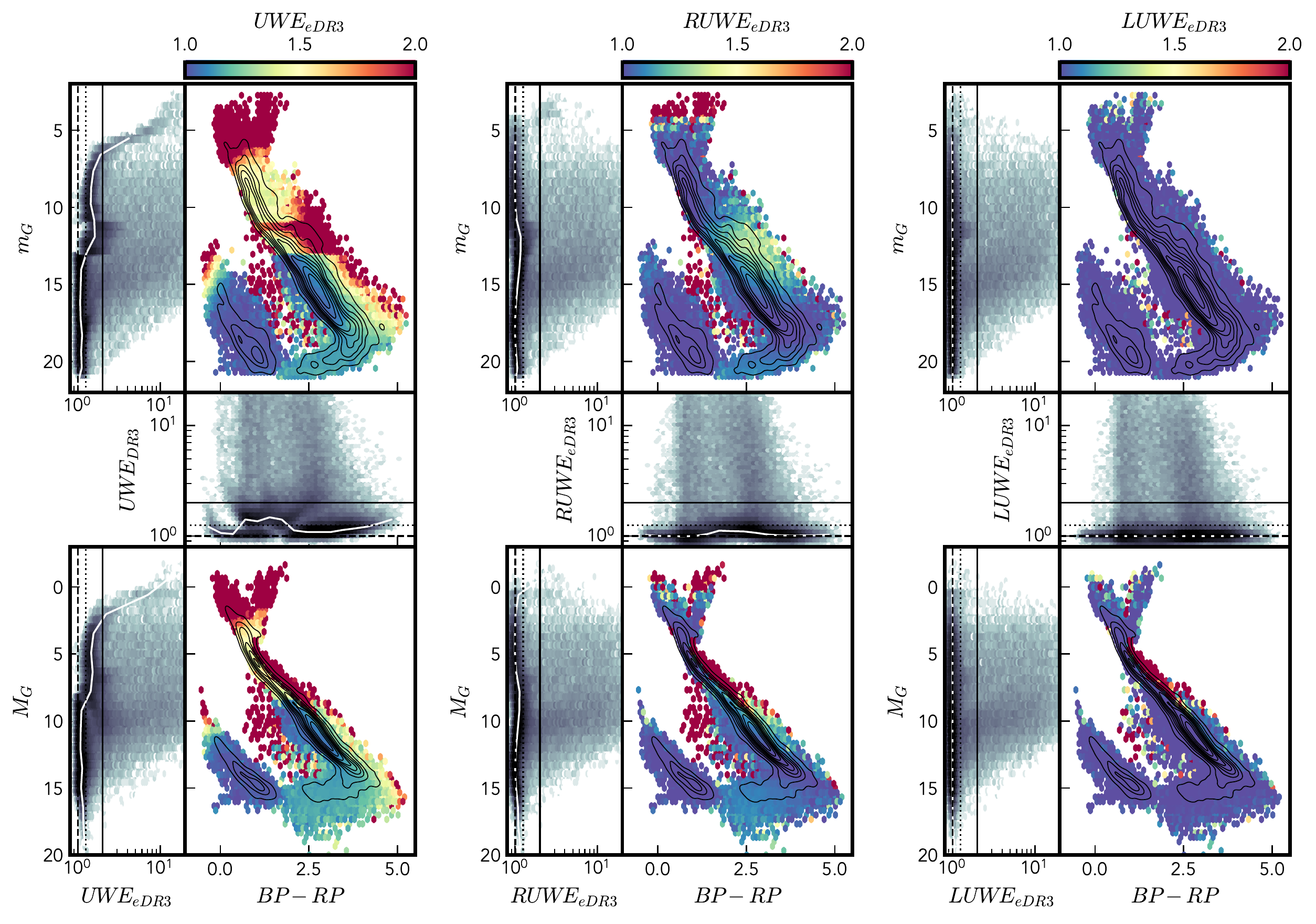}
\caption{Coloured panels: The distribution of the 41\textsuperscript{st} percentile of Unit Weight Error ($UWE$, left), Renormalised Unit Weight Error ($RUWE$, middle) and Local Unit Weight Error ($LUWE$, right) as a function of position on the HR diagram. Thin black lines are contours showing the underlying density distribution (see Figure \ref{hrSimple} for more details). Grey-scale panels: distribution of each error as a function of apparent magnitude ($m_G$, upper), colour ($BP-RP$, middle) and absolute magnitude ($M_G$, lower) coloured by number density (on a log-scale). The 41\textsuperscript{st} percentile error is shown by a white line, and lines representing errors of 1, 1.25 and 2 are shown as dashed, dotted and solid black lines respectively.}
\label{compweighterror}
\end{figure*}

\begin{figure*}
\centering
\includegraphics[width=0.98\textwidth]{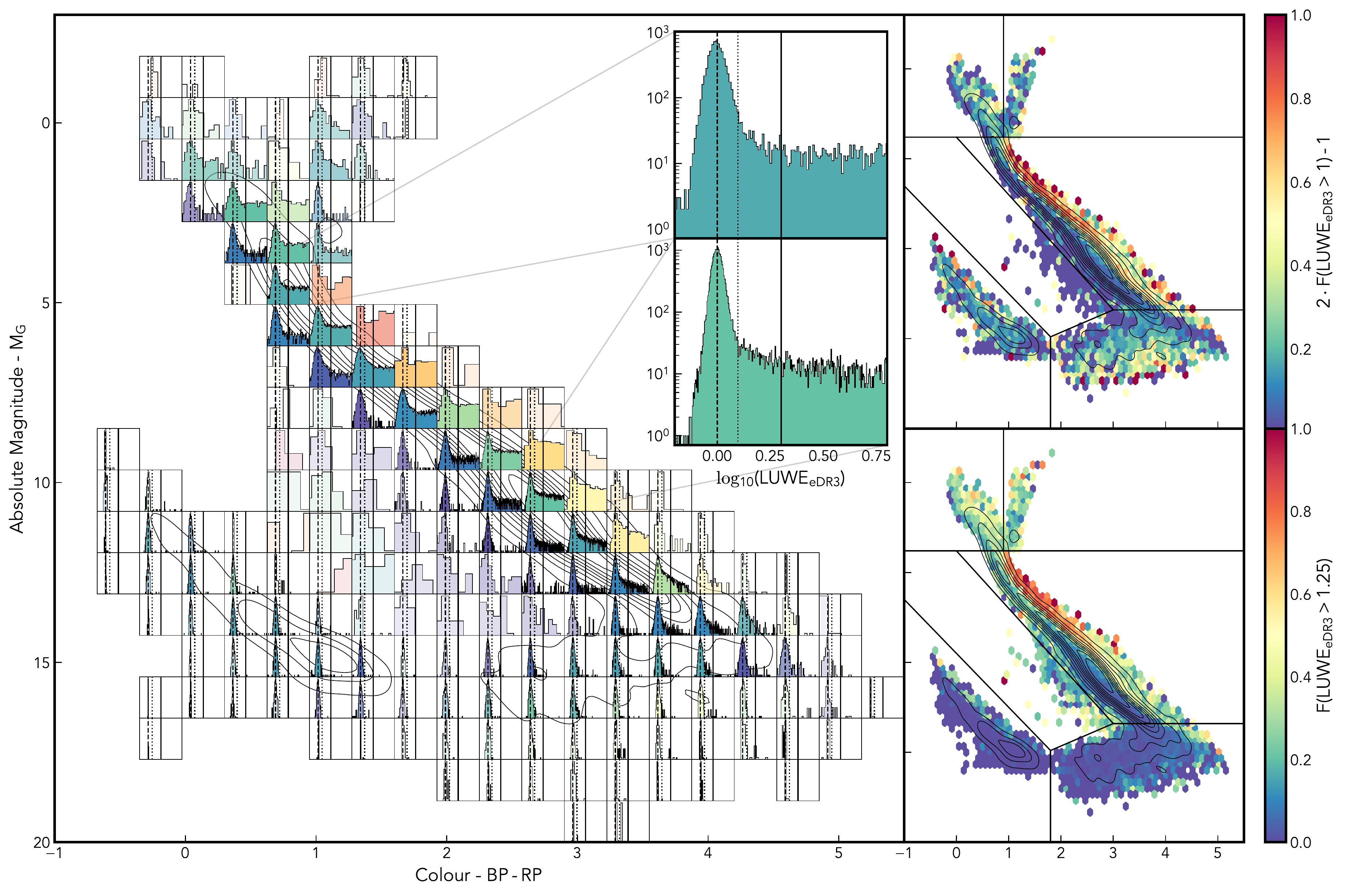}
\caption{The equivalent of Figure \ref{hrRuweHists} but now using our Local Unit Weight Error (LUWE) instead of RUWE.}
\label{hrLuweHists}
\end{figure*}

\subsubsection{Renormalising the UWE/RUWE}

The calculation of Unit Weight Error (UWE) requires normalising the residuals between observations and a single source model using the astrometric error, $\sigma_{ast}$. This is not a known quantity, depending at the very least on the colour and apparent magnitude ($m_G$) of the source, which affects the performance of the telescopes' CCDs. Incorrectly estimating $\sigma_{ast}$ will cause the peak in $UWE$ of a population dominated by single stars to stray from the expected value of 1.

Phrased another way, the modal UWE of a population of well behaved sources of a particular magnitude and colour can be used to infer the true astrometric error, showing us (if we could identify a truly well-behaved population) how to rescale whatever error was previously assumed.

A simple formulation of the expected magnitude dependence of $\sigma_{ast}$ is used when calculating $UWE$. This fails to capture all the complexity of the instruments sensitivity, some regions (most notably $13>m_G>11$) have $UWE$ significantly greater than 1.

This is why the renormalization step must be performed, as detailed in \citet{Lindegren18}, shifting the UWEs of all stars of a given colour and apparent magnitude by an empirically derived factor, to give renormalised values (RUWE). This is done by rescaling the whole sample, as a function of colour ($BP-RP$) and apparent magnitude, such that the 41\textsuperscript{st} percentile of $RUWE \rightarrow 1$. The choice of the 41\textsuperscript{st} percentile represents the fact that we should expect the median single star to have $RUWE=1$ but that any sample will also contain contamination by binary systems and other sources of noise (which in general increase noise) such that the median of the whole sample should have $RUWE>1$. The fixed percentile is a pragmatic approximation presented in \citet{Lindegren18} which we'll adopt here also.

Examining the distribution of $RUWE$, as shown in the middle column in Figure \ref{hrRuweHists}, we see that there are some parts of the HR diagram where the GCNS sample of sources is biased to high RUWEs, with distributions peaking above 1. Main sequence stars with $15\gtrsim m_G \gtrsim 11$ are particularly badly affected. A high fraction of sources with $RUWE>1$ might be seen as evidence for a large binary fraction. However, this should be correlated with a high fraction of sources with $RUWE>1.25$ (the most extreme values that could be expected from stochastic scatter of single stars). Examining the righthand panels of Figure \ref{hrRuweHists}, it is clear that in some regions this is not the case - RUWE is high not because of companions but due to imperfect renormalization for this sample.

We believe the reason that the renormalization of the sample, which performs well for the total catalogue, fails for this small local subset is that any such statistic calculated across apparent magnitude and colour space is dominated by stars at distances well-beyond our sample. These stars, which must be intrinsically brighter, may have a different $\sigma_{ast}$ because of their different characteristics such as variations in their spectra, amount of extinction, binary fraction (for a fixed apparent magnitude), or because of an unmodelled scaling of $\sigma_{ast}$ with distance. 

Regardless of the reason, it is relatively simple to correct for this effect - by rescaling the RUWEs\footnote{Note that this process is broadly equivalent to rescaling the $UWE$ directly, we work in $RUWE$ only because it is the more readily accessible Gaia data product.} over the GCNS sample, in an equivalent manner as the original rescaling of UWEs to produce RUWEs. To do this, we calculate the $41_{st}$ percentile of $RUWE$ on a uniform $100 \times 100$ grid spanning $2<m_G<22$ \& $-1<BP-RP<5.5$ and then linearly interpolate to find the correction factor based on the apparent magnitude and colour of each source. Where there is insufficient data (less than 3 sources in a given bin), we assume a median $RUWE$ of 1.

We do this for all eDR3 sources in GCNS, and their equivalent in DR2, producing a new statistic which we'll call the Local Unit Weight Error (LUWE) as can be seen in Figure \ref{compweighterror}. Here we see that the UWE\footnote{Calculated via \textit{Gaia}'s  $\chi^2=$\texttt{ASTROMETRIC\_CHI2\_AL}, $N_{o}=$\texttt{ASTROMETRIC\_N\_GOOD\_OBS\_AL} and $N_{p}$ is the number of parameters used in the solution (usually 5 or 6, recorded by \texttt{ASTROMETRIC\_PARAMS\_SOLVED}), as $UWE=\frac{\chi^2}{N_{o} -N_{p}}$.} clearly has large deviations from the expected modal value of 1, and whilst they are much reduced some of this effect is still visible in $RUWE$. $LUWE$ shows much more uniform and expected behaviour, struggling only in regions with very few sources.

Figure~\ref{hrLuweHists} shows the equivalent of Figure~\ref{hrRuweHists} using LUWE. The improvement is clear -- now the whole main-sequence has $LUWE$ peaking at 1, deviating only for extremely bright sources which we expect to be dominated by photometric binaries. There is now a much cleaner correlation between $F(LUWE>1)$ \& $F(LUWE>1.25)$. There still seems to be an offset for the lowest luminosity sources but this is a sample we will not explore in detail in this work as Gaia struggles to reliably measure these very dim objects.

It is interesting to note that the correction is performed in terms of apparent magnitude (ignoring extinction), but here we are showing the success in absolute magnitude. This is due to the fact that the error itself depends primarily on apparent magnitude, and once the correction is performed it translates into $M_G$.

Examining the histograms, we can get a feel for the number and character of possible unresolved binary systems. Throughout the main sequence there is a significant continuum of high LUWE sources. They increase in number as we trace along the MS towards more massive stars, and across it towards particularly bright stars of a given colour. At the extremes of the colour-magnitude diagram, the high LUWE component can dominate over the peak consistent with single sources (which is also why we cannot perform the normalisation in $M_G$ space, as these would be erroneously shifted back to 1). In comparison, the high LUWE sources amongst the white dwarfs are much rarer, too few to form a continuum, but again more prevalent for brighter sources of a fixed colour.

We anticipate that for larger subsamples the difference between $RUWE$ and a calculated $LUWE$ will decrease - but may be important even at significantly larger distances. We might also expect the correction factor to vary over a given sample, especially if it contains sources at distances differing by orders of magnitude. Thus, in some cases, it may have to be performed as a function of distance as well. Given our local sample, and the relatively small number of sources (compared to the full Gaia dataset), we do not attempt such a distance dependent correction here.

We will use our subsample selected by the above quality cuts, and the $LUWE$ statistic throughout this work.

\section{Astrometric deviations}
\label{astrometricdeviations}

\begin{figure*}
\centering
\includegraphics[width=0.98\textwidth]{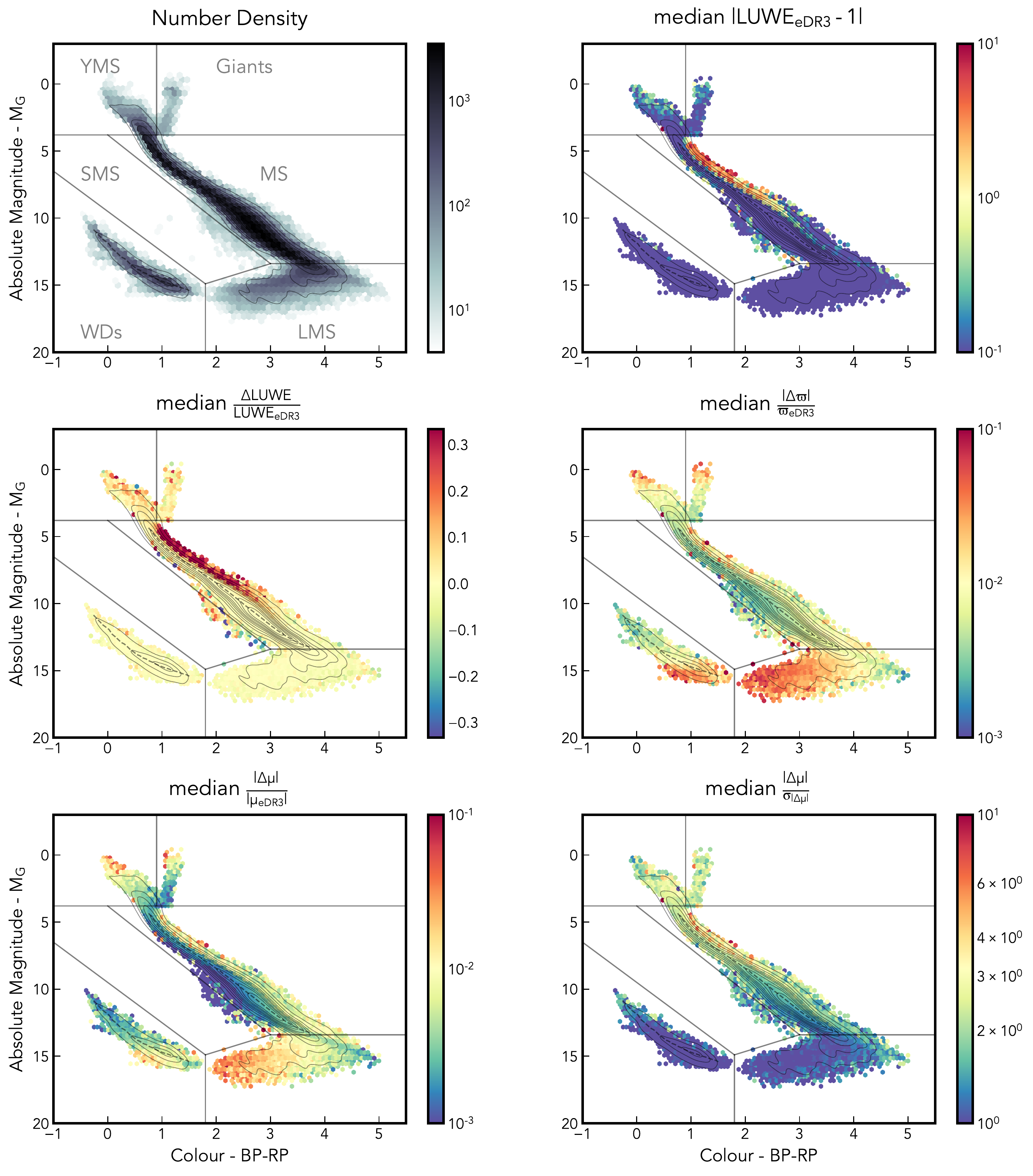}
\caption{Hertzsprung Russel diagram for sources passing our quality cuts in the Gaia catalogue of Nearby Stars (within 100 pc), similar to Figure \ref{hrSimple}. As well as the number density (top left) we also show the variation with LUWE (top right), change in LUWE (middle left), change in parallax (middle right) and magnitude of proper motion anomaly normalised by both the observed proper motion (bottom right) and the inferred error (bottom left), all between Gaia DR2 and eDR3.}
\label{hrRuwes3}
\end{figure*}

\begin{figure}
\centering
\includegraphics[width=0.48\textwidth]{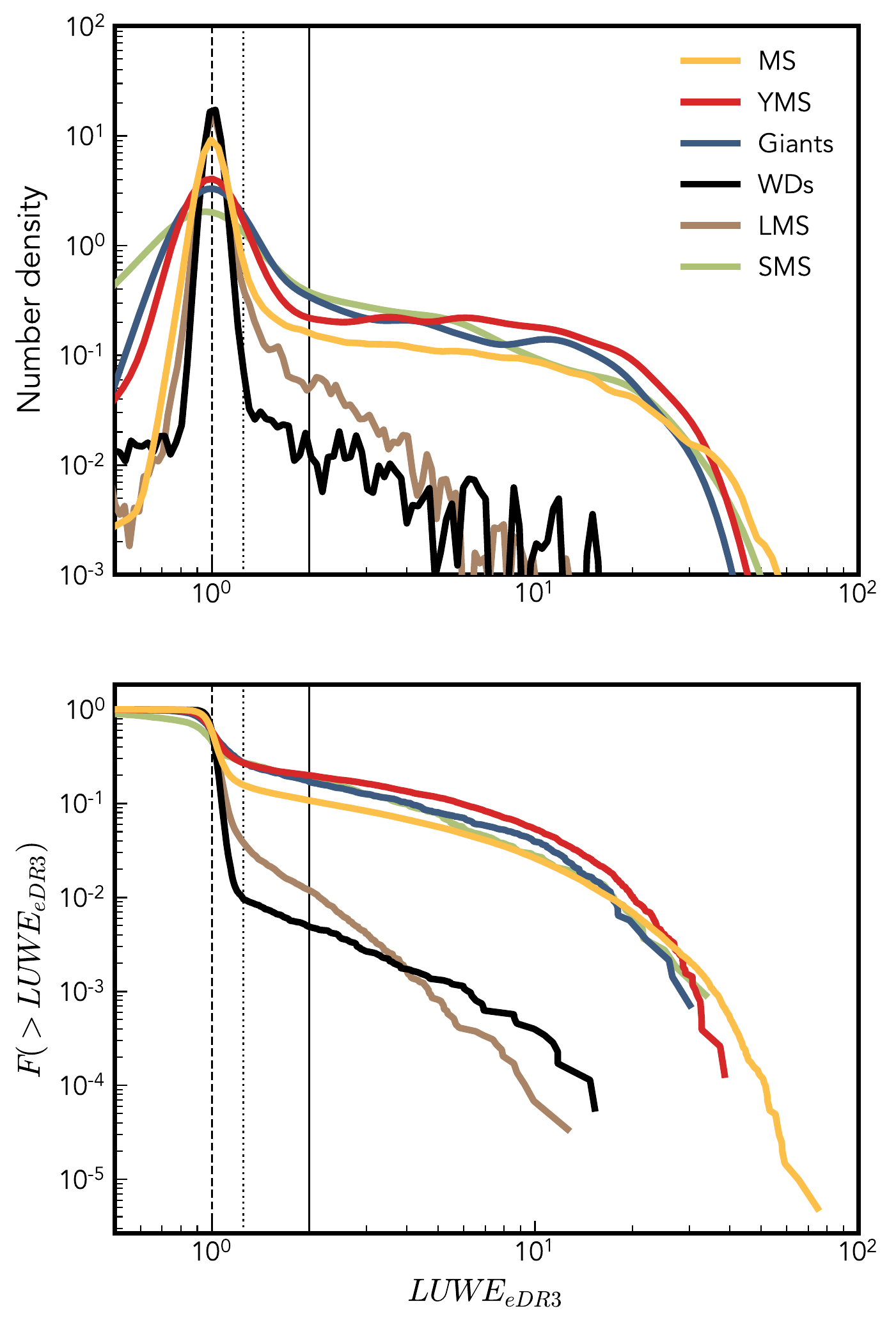}
\caption{The distribution of LUWEs separated by classification from the HR diagram. The dashed, dotted and solid black vertical lines show LUWEs of 1, 1.25 and 2 respectively. In the top panel we show the histogram of all sources (with an arbitrary normalization) and in the bottom panel the fraction of sources of each type with a $LUWE$ greater than a given value.}
\label{luwehist}
\end{figure}

\begin{figure*}
\centering
\includegraphics[width=0.98\textwidth]{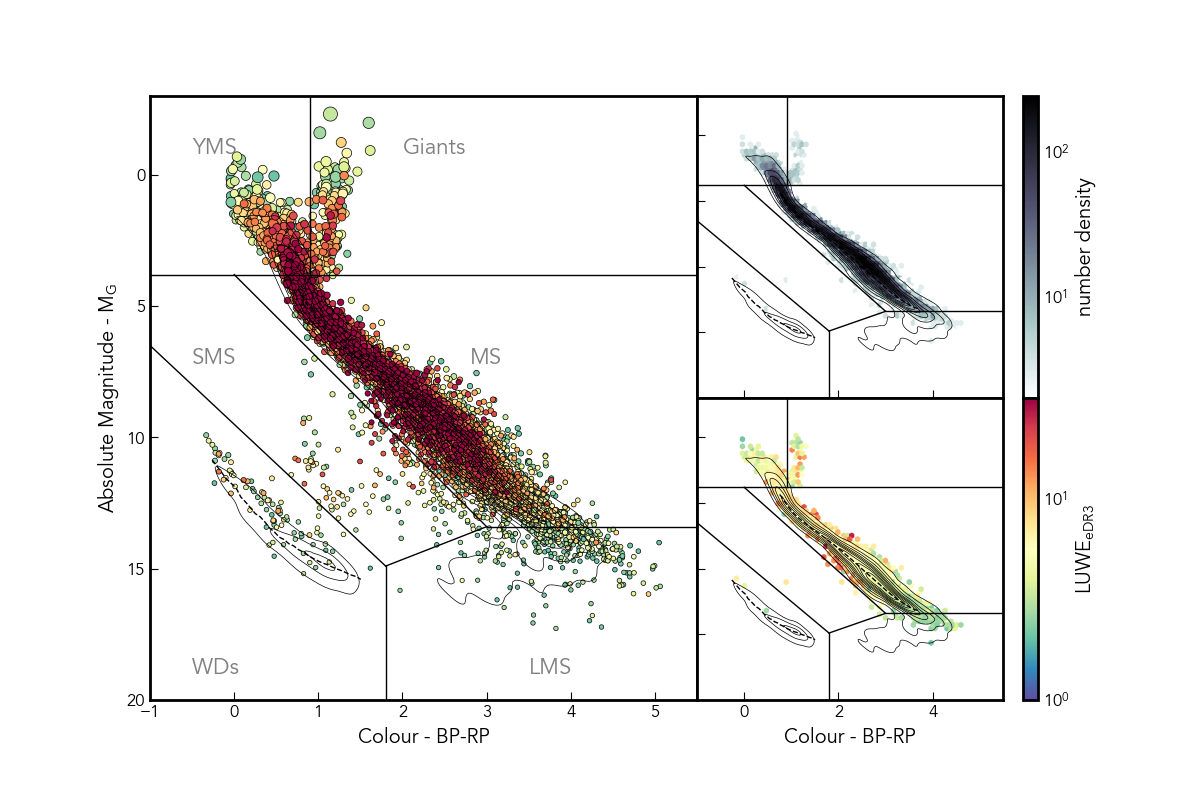}
\caption{22,699 candidate binary systems in the Gaia Catalogue of Nearby Stars. In the left hand panel we plot each individual system, coloured and sorted by $LUWE$ and with a size set by the apparent magnitude. In the right hand panels, we show the distribution of these sources and their median $LUWE$.}
\label{gaia_binaryHR}
\end{figure*}

\begin{figure}
\centering
\includegraphics[width=0.49\textwidth]{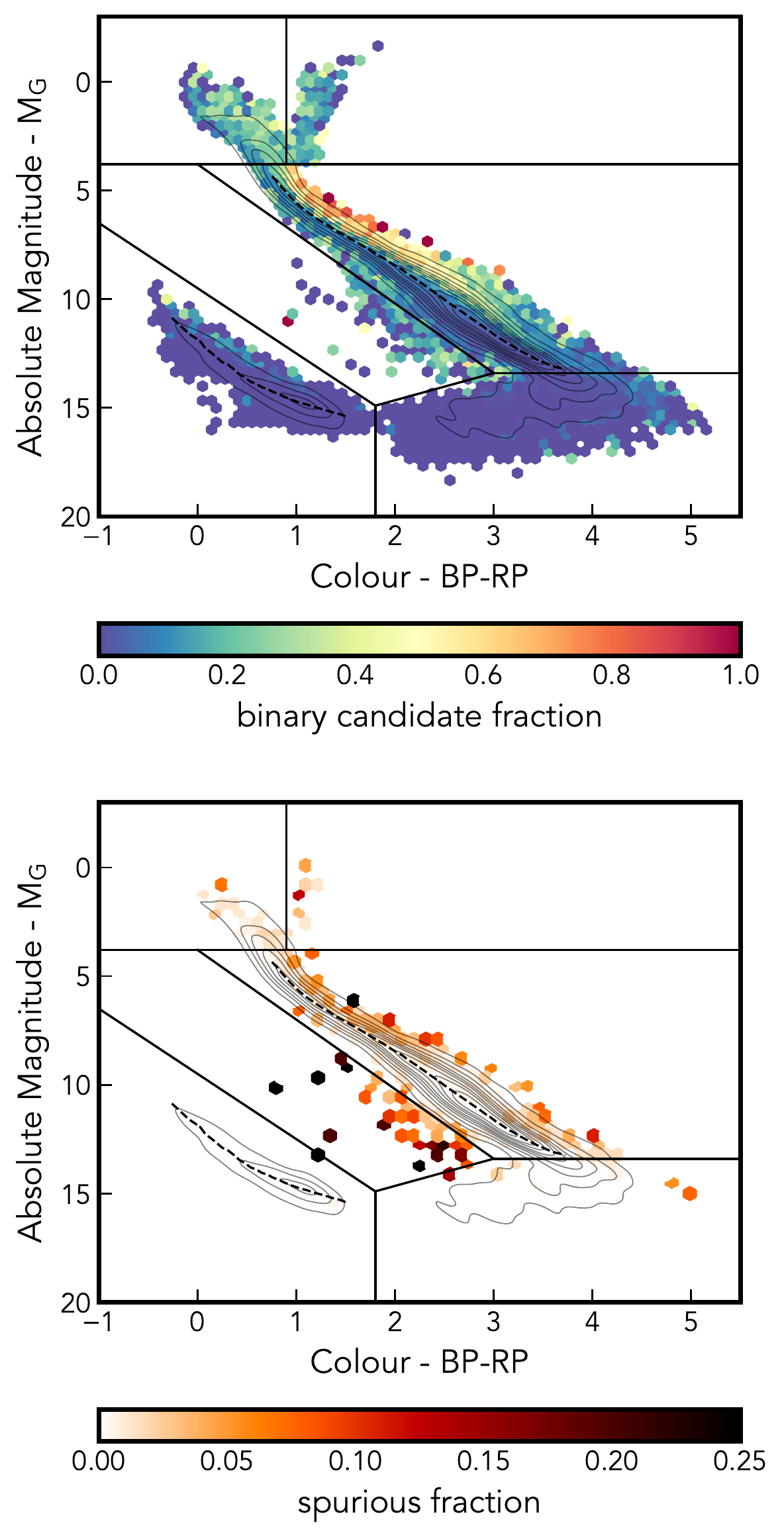}
\caption{The top panel shows the fraction of binary sources compared to all sources in GCNS, as a function of position on the HR diagram. In the bottom panel we reverse the condition that reliable binaries should have $\Delta LUWE > -\frac{LUWE}{3}$ to estimate the degree of contamination by spurious sources with high $LUWE$ incompatible with the expectation of binary motion. Here we include bins only if they contain 3 or more sources, hence not all sources visible in Figure \ref{gaia_binaryHR} are included.}
\label{gaia_binfrac}
\end{figure}

\begin{figure}
\centering
\includegraphics[width=0.49\textwidth]{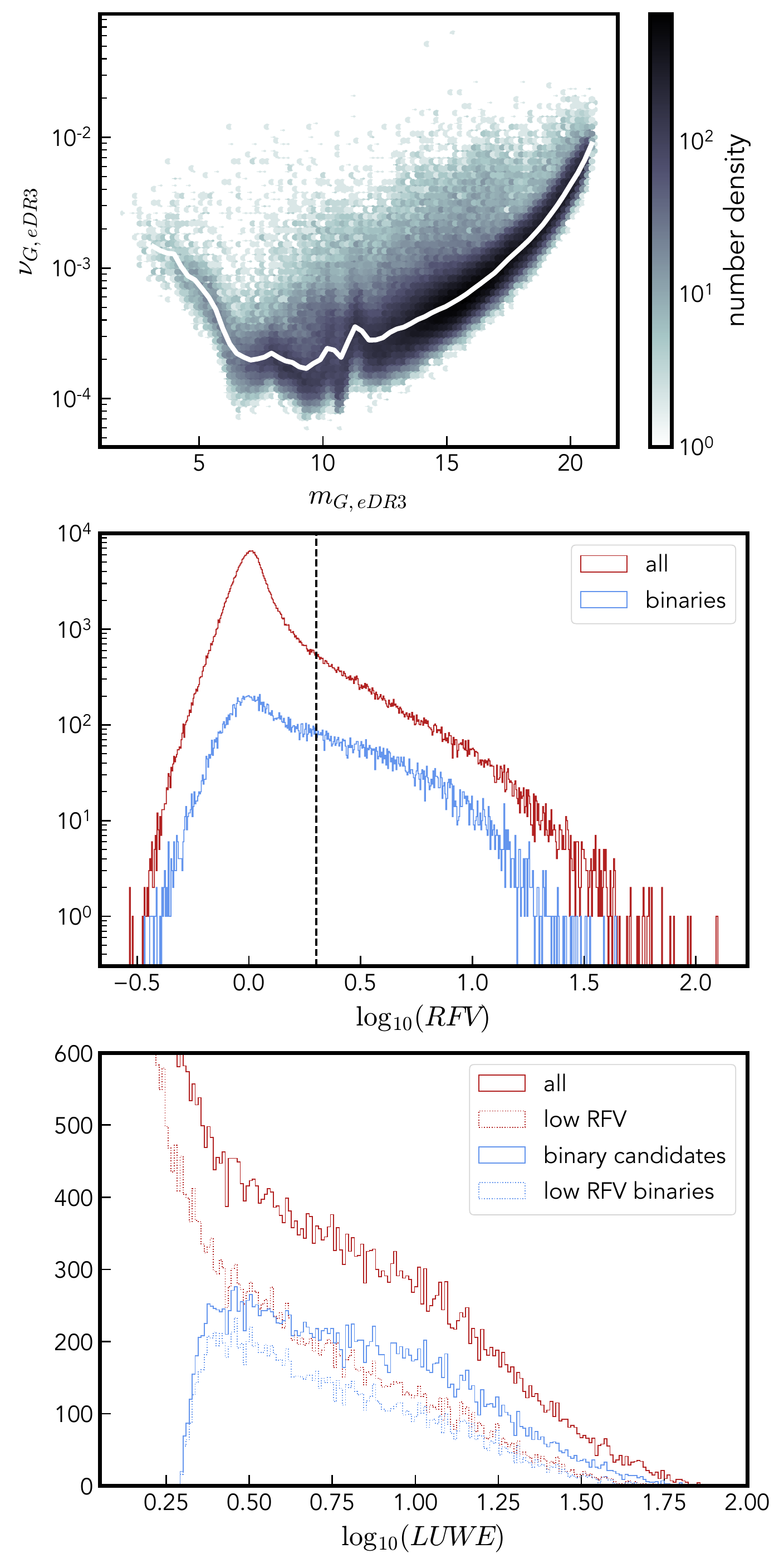}
\caption{In the top panel we show the distribution of flux variability (equation \ref{fluxvareq}), showing the number density of sources and the median, as a function of magnitude as a thick white line. The middle panel shows the distribution of Relative Flux Variability (RFV), equal to the variability of each source normalised by the median for sources of the same magnitude. In the middle plot we also a dashed vertical line at $RFV=2$, the suggested boundary between high and low variability stars. In the bottom panel we show the distribution of $LUWE$ as separated by $RFV$ for single stars and binary candidates.}
\label{rfvhistogram}
\end{figure}

\begin{figure}
\centering
\includegraphics[width=0.49\textwidth]{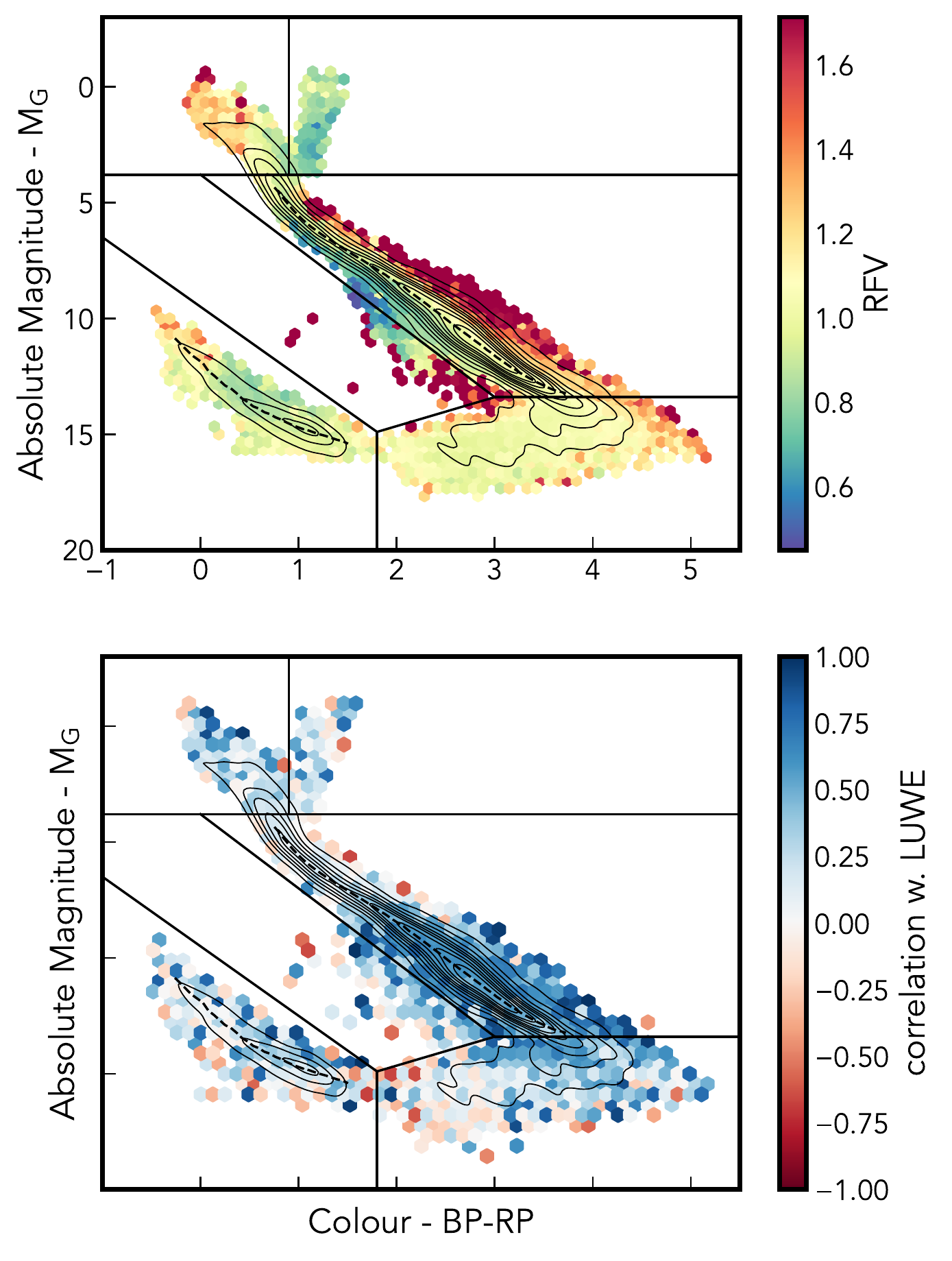}
\caption{Median Relative Flux Variability (RFV) as a function of position on the HR digram (top) and the correlation coefficient between RVF and LUWE in each hexagonal bin (bottom).}
\label{rfvcorrelation}
\end{figure}

In Figure~\ref{hrRuwes3}, we show various astrometric and photometric properties as a function of position on the HR diagram. As observed in B+20 we can see, in the $LUWE$, a clear multiple star main sequence (above the median MS brightness) with high LUWE, as well as an excess of LUWE for the brightest YMS and Giant stars. These populations also have large positive $\Delta LUWE$, which follows from arguments in P+21 that these are highly correlated, especially for binaries with periods close to the length of the observing window. Given that we are still probing a range of periods here and we expect longer period ($\gtrsim 3$ years) binaries to be more common \citep{Raghavan10}, it is not surprising that this effect is so apparent.

As discussed in P+21, $\Delta LUWE$ is also very instructive as to the nature of the SMS population -- which often have smaller $LUWE$ in eDR3 than DR2, suggesting that a significant fraction of these sources are spurious solutions. These may be caused by a few erroneous astrometric measurements, whose impacts are reduced by a longer observing period. Not all sources in this region have $\Delta LUWE<0$, suggesting there is a population of accurately measured systems present here, likely corresponding to MS+WD binaries (see for example \citealt{Ren18,RM21} for a much fuller analysis of these objects). It is in this region that cleaning the sample with cuts on $\Delta LUWE$ is thus particularly important for identifying true systems of interest.

Parallax anomaly,
\begin{equation}
\Delta \varpi= \varpi_{eDR3} - \varpi_{DR2},
\end{equation}
is small, rarely more than 10\% (P+21, remember that we have cut sources with $\frac{\varpi}{\sigma_\varpi}<10$ and this will effectively limit observable parallax shifts as well). It is significant for the LMS, but also shows some signs of being increased for populations associated with other astrometric error (the bright binary MS) and the brightest evolved stars. As argued in P+21 we expect some systems to have both large astrometric errors and parallax shifts due to binarity. In general the latter will be rarer than the former, as it is only significant for a very specific part of parameter space (periods harmonic with 1 year, low eccentricity and particular orientations).

We also see significant proper motion anomaly,
\begin{equation}
\Delta \bm{\mu}= \bm{\mu}_{eDR3} - \bm{\mu}_{DR2},
\end{equation}
for the systems thought to be binaries. We show the relative change in proper motion, and the significance of the change, calculated via
\begin{equation}
\sigma_{|\Delta \bm{\mu}|}=\frac{\sqrt{\Delta \mu_{\alpha^*}^2 (\sigma_{\mu_{\alpha^*,eDR3}}^2 + \sigma_{\mu_{\alpha^*,DR2}}^2) + \Delta \mu_{\delta}^2 (\sigma_{\mu_{\delta,eDR3}}^2 + \sigma_{\mu_{\delta,DR2}}^2)}}{|\Delta \bm{\mu}|}
\end{equation}
where $\alpha^*$ and $\delta$ denote the east and north directions respectively. The behaviour of these measurements mostly replicates the distribution of LUWE. Large $|\Delta \varpi|$ and $|\Delta \bm{\mu}|$ are also seen for some of the brightest and dimmest sources, suggesting that these extreme magnitude sources may be suffering from some systematic effect.


Figure~\ref{luwehist} shows the distribution of $LUWE$ split between the different classes of object as determined from the HR diagram. 

We can gauge the natural spread of $LUWE$s from the distribution of sources with $LUWE<1$: Bright stars (YMS and Giants) and SMS stars have a much broader spread, while MS stars, WDs and LMS stars are much tighter. Moving to higher $LUWE$, we see the distribution of WDs and LMS stars drops off quickly, suggesting only a small fraction of these sources are binary systems within the range of visible periods.

There seem to be many more binaries among more massive stars, 10\% of MS stars have $LUWE>2$, and that number rises to $\sim$20\% for YMS, Giant SMS sources. In comparison, around 1\% of LMS stars have high $LUWE$, and only 0.5\% of WDs.

\section{A catalogue of nearby unresolved binaries}
\label{binarycandidates}

We can produce a list of nearby sources which are likely to host at least one companion. We use a simple criteria for a likely unresolved binary candidate of
\begin{itemize}
\item $LUWE_{eDR3}>2$
\item $\Delta LUWE > -\frac{LUWE}{3}$
\end{itemize}
($\Delta LUWE = LUWE_{eDR3}-LUWE_{DR2}$) on sources already meeting our other general quality cuts.

We find 22,699 systems within the GCNS sample, a small subsample of which can be seen in table \ref{binarydata}. These sources are shown on the HR diagram in Figure~\ref{gaia_binaryHR}. We see that the whole HR diagram is spanned by candidate binary systems, with a high number especially in the MS, YMS and Giant regions.

The criterion of $LUWE>2$ is stronger than the cutoff of 1.25 suggested in P+21. Real data, with the extra sources of noise such as blending and other unmodeled noise, can be expected to have a wider spread than the idealised simulations presented there, and in Figure \ref{luwehist} we can see that this is particularly true for giant stars. Choosing a larger $LUWE$ cutoff does not significantly reduce the number of binary candidates, as can be seen by the shallow slopes in the cumulative distributions (with the partial exception of LMS stars) and gives us more confidence that the contamination by single sources is small. In general we would advise choosing cutoff criteria for binaries/single stars by direct reference to the distribution of $LUWE$ for stars in your sample, rather than using a particular literature value.

Though there are relatively few WD candidates, it is interesting to note that the majority sit at or above a magnitude approximately 0.75 above the median for all white dwarfs. This is the offset expected from two sources of identical luminosity, suggesting that WD binaries may have a predisposition to being similar in mass and magnitude (though sufficiently different that the center of light is still offset from the center of mass).

The full catalogue of candidates can be accessed at \url{https://zenodo.org/record/6053827}.

\subsection{Binary candidate fraction}
\label{bincanfrac}

In Figure \ref{gaia_binfrac}, we show the fraction of systems selected by the above criteria as a function of colour and absolute magnitude. The fraction is low along the median track for the Main Sequence. The fraction increases rapidly for brighter systems, which we expect to be dominated by photometric binaries and higher multiplicity systems. It also increases for particularly dim/blue MS sources, potentially caused by binaries containing a MS star and a WD, causing a significantly bluer system without being significantly brighter.

There is also a clear variation in binary fraction along the MS -- which we can ascribe to a combination of two effects, one physical and the other observational. We do expect more massive stars (for which absolute magnitude is a close proxy) to have a higher binary fraction \citep{Duchene13,Offner20} and the trend along the MS agrees with this. However, there is a second effect based on our ability to resolve significant astrometric deviations: only systems with periods $\lesssim 10$ years will have significant $LUWE$. As we move to lower mass dimmer systems, this requires the orbit to be smaller, and thus the astrometric signal becomes more dominated by noise.

Sources on the YMS have a higher fraction generally, again showing a gradient of increasing binarity moving from dimmer sources to the brightest. The Giants are too few in number to infer any particular trend, but also show a higher apparent binary fraction. For both of these populations, particularly the Giants, there is a possibility that the high $LUWE$ may be caused not by a companion but by variations in the brightness across the surface (and thus the position of the photocentre) resolvable on sufficiently large and close stars \citep{Pasquato11}.

Again we see some evidence of a binary WD population, significantly brighter than the main population of WDs and with a higher candidate fraction.

We can also estimate the degree of contamination by spurious high $LUWE$ sources, by looking at systems which pass all of our other criteria but have large negative $\Delta LUWE$, which cannot be reconciled with binary motion (as discussed in detail in P+21). There is a small amount of contamination in regions with high binary fractions, but much less than binary fraction itself. The only exception is the SMS, where the spurious fraction is of similar magnitude to the binary fraction, suggesting that the a significant number of sources in this region may have high $LUWE$ caused by some other property than binarity. It is worth noting that this region is the most affected by cutting out sources in dense environments, and is where we would expect distant objects with erroneously high parallaxes to reside. This doesn't mean that all SMS binary candidates are invalid, but that we should employ the highest degree of scepticism for this population.

Given the sensitivity of astrometric measurements and the expected distribution of binary periods \citep{Raghavan10} we would expect a sample composed of entirely binaries to have an astrometric binary candidate fraction of $\sim 20-30\%$. This is what we see, for example, for YMS and Giant sources. However there are some regions of the HR diagram where the observed fraction significantly exceeds this nominal upper limit, most notably at the bright end of the MS. We suggest two possible explanations for this quandry, both of which would produce the observed effect but are hard to unpick from each other (and both may be contributing):
\begin{itemize}
\item Multiple stars - Systems with 3 or more bodies must, in order to be approximately stable over long periods, contain orbits with significantly different periods. Thus the chance of any one pair in the multiple having an orbital period relevant for astrometry is increased. Similarly the presence of more luminous sources leads to a higher magnitude, in good agreement with where these are observed on the HR diagram.
\item Biased parallaxes - If these systems contain binaries we can expect the parallax to be biased to some degree. Especially for binaries with periods close to 1 year this shift can be significant, if $\Delta \varpi$ is negative this will lead to an overestimate of the inferred luminosity of the source. Any binary with a significant parallax shift will also likely have a high $LUWE$ and thus a population of astrometric candidates will be dispersed above and below the binary main sequence (with those falling below being an insignificant contribution to this relatively dense region of the HR diagram).
\end{itemize}

\subsubsection{Variability}

If the light from a source is spread over a large angular diameter (compared to the astrometric error), as is the case for binaries and some giant stars, then variability can cause the centre of light to shift. These systems are often called Variability Induced Movers (VIMs, \citealt{Wielen96}).

Perhaps the easiest system in which to picture this is a long period (unresolved) binary system, with a varying light ratio $l$, perhaps due to starspots or a transit. This will cause the position of the centre of light, relative to the centre of mass, to vary over time without requiring any binary motion. Single giant stars can also show this effect, if the star itself is enlarged sufficiently and the surface brightness is anisotropic and time-varying. 

VIM is another potential cause of high astrometric error -- in most cases still associated with a binary. We can gauge this effect in the Gaia data by comparing the level of variability of stars to some measure of astrometric noise.

Gaia records the mean and standard deviation of the flux measurements it takes from each source. We will use the G band measurements to find a flux variability
\begin{equation}
\label{fluxvareq}
\nu_G=\frac{\sqrt{N_{obs,G}}\sigma_{F_G}}{F_G},
\end{equation}
(the reciprocal of \textit{Gaia's} \texttt{PHOT\_G\_MEAN\_FLUX\_OVER\_ERROR} multiplied by the square root of \textit{Gaia's} \texttt{PHOT\_G\_N\_OBS})
and normalise based on the median relative error, $\langle \nu_G \rangle (m_G)$ for sources of similar magnitude giving a relative flux variability (RFV) calulated as
\begin{equation}
RFV=\frac{\nu_G}{\langle \nu_G\rangle(m_G)}.
\end{equation}
Much like the $UWE$ this is expected to peak at one, and particularly variable sources should have $RFV>1$.

In Figure~\ref{rfvhistogram}, we show the distribution of RFV and how it relates to the LUWE distribution and our subset of binary candidates. 35\% of our candidates have $RFV>2$ whilst only 21\% of all GCNS stars do, suggesting either that some of our candidates may show evidence of being VIMs or that variable stars form a (minor) contamination of our sample.

In Figure~\ref{rfvcorrelation}, we show the variation of $RFV$ across the HR diagram. We calculate the Pearson correlation coefficient between RFV and LUWE - which will tend to 1 if flux variability and astrometric noise are positively correlated, and 0 for no correlation. The sources with the highest intrinsic flux variability are the YMS and the photometric binary MS (above the median MS). Across much of the HR diagram flux variability is highly correlated with astrometric noise. The only region with clearly no correlation is the bright MS photometric binaries, some of the sources for which purely astrometric motion is easiest to resolve. The high correlation amongst dimmer MS stars may be evidence that there are significant numbers of VIMs in this region (in binaries that would otherwise be undetectable astrometrically, for example, if the period was too long for the survey). 

Given that variability induced motion is most visible in binaries of longer periods (with wider but still unresolved separations), closer to the peak of the expected binary period distribution, it seems plausible that these comprise a significant fraction of the astrometric deviations for this sample.

Another cause of both variability and astrometric uncertainty is a partially resolved binary which is registered as a combined single source when scanned at some angles and two separate (or just the brighter) at other angles. These partially resolved binaries, and their indicators in \textit{Gaia}, are discussed in more detail in Appendix \ref{IPDbinary}.

Equally, this could be a tell-tale that there is another spurious population in our catalogue where high variability and high LUWE are both signs of a poorly described source. These sources are the minority of binary candidates, and thus we do not exclude them here, but do include RFV in our catalogue such that it's easy to separate them in future.


\subsection{Behaviour of MS sources}

For the final part of this paper, we examine the Main Sequence sources and unresolved binary candidates in more detail.

\begin{figure}
\centering
\includegraphics[width=0.49\textwidth]{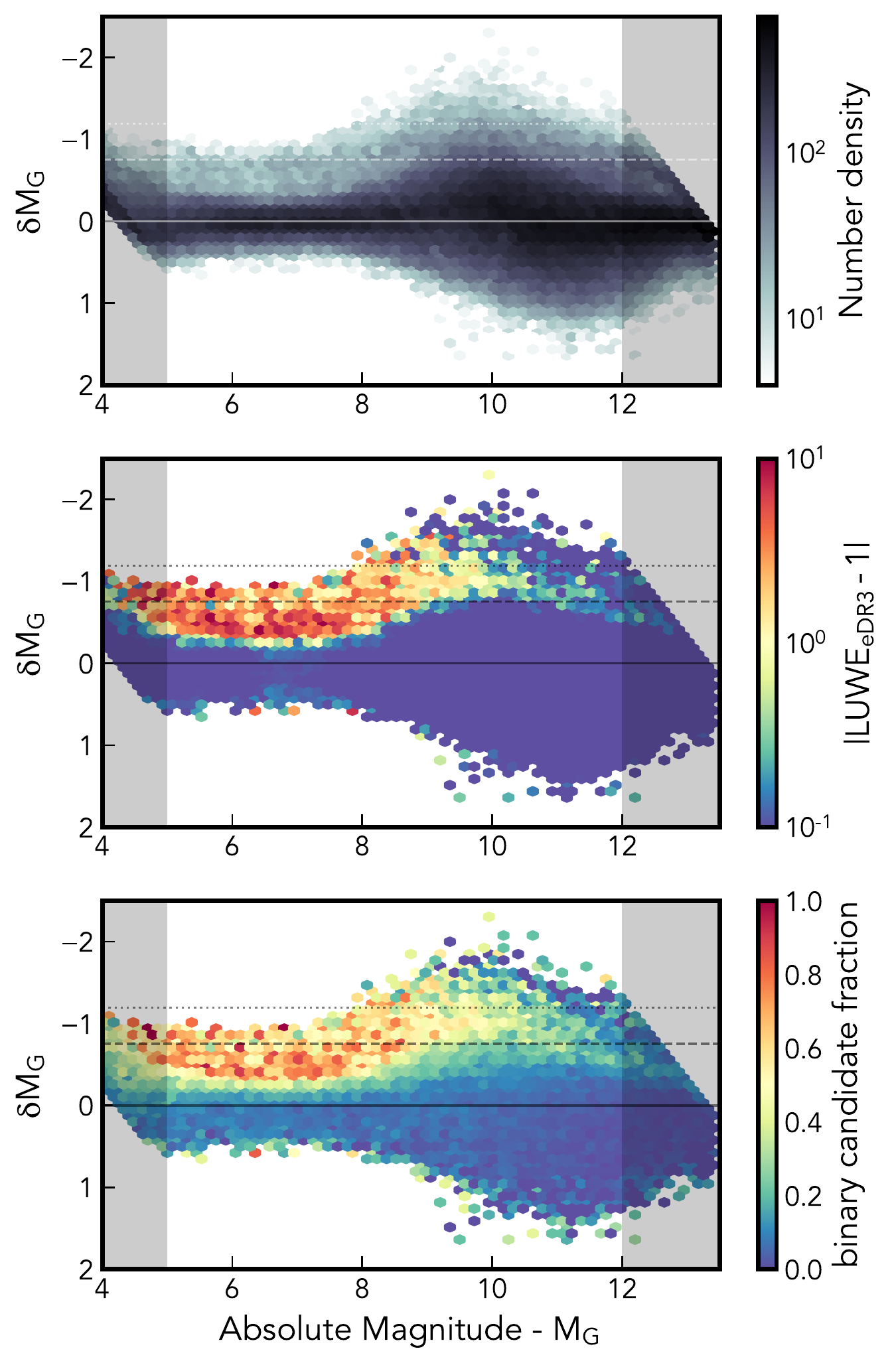}
\caption{The variation in magnitude of Main sequence stars from the median (as a function of colour) against absolute magnitude. Shaded regions are those for which the selection criteria for MS stars (as shown in Figure \ref{hrSimple}) cuts off some bright/dim sources and the median is likely to be biased. We also show as horizontal solid, dashed and dotted lines a magnitude shift of $0$, $-\frac{5}{2}\log_{10}(2)$ and $-\frac{5}{2}\log_{10}(3)$ respectively. These represent the largest possible shifts for 1, 2 and 3 stars respectively (for which $\delta M_G=-\frac{5}{2}\log_{10}(N)$ for N identical luminosity stars).}
\label{msCutout}
\end{figure}

In Figure \ref{msCutout} we zoom in on, and flatten, the MS as shown in Figure \ref{hrRuwes3}. Again we see significant $LUWE$ for stars brighter (negative $\delta M_G$) than the median, where we expect MS binaries to reside. This is particularly pronounced for brighter stars. 

For stars dimmer than $\sim$10\textsuperscript{th} magnitude we see the highest LUWEs (and binary fraction) at some intermediate $\delta M_G$. We know that identical stars in a binary should have no offset between the centre of light and the centre of mass and that may be why these systems have smaller LUWEs. That this behaviour is not seen at higher magnitudes is likely due to contamination, either by triples or binaries with highly biased parallaxes, as discussed in section \ref{bincanfrac}.

\begin{figure}
\centering
\includegraphics[width=0.49\textwidth]{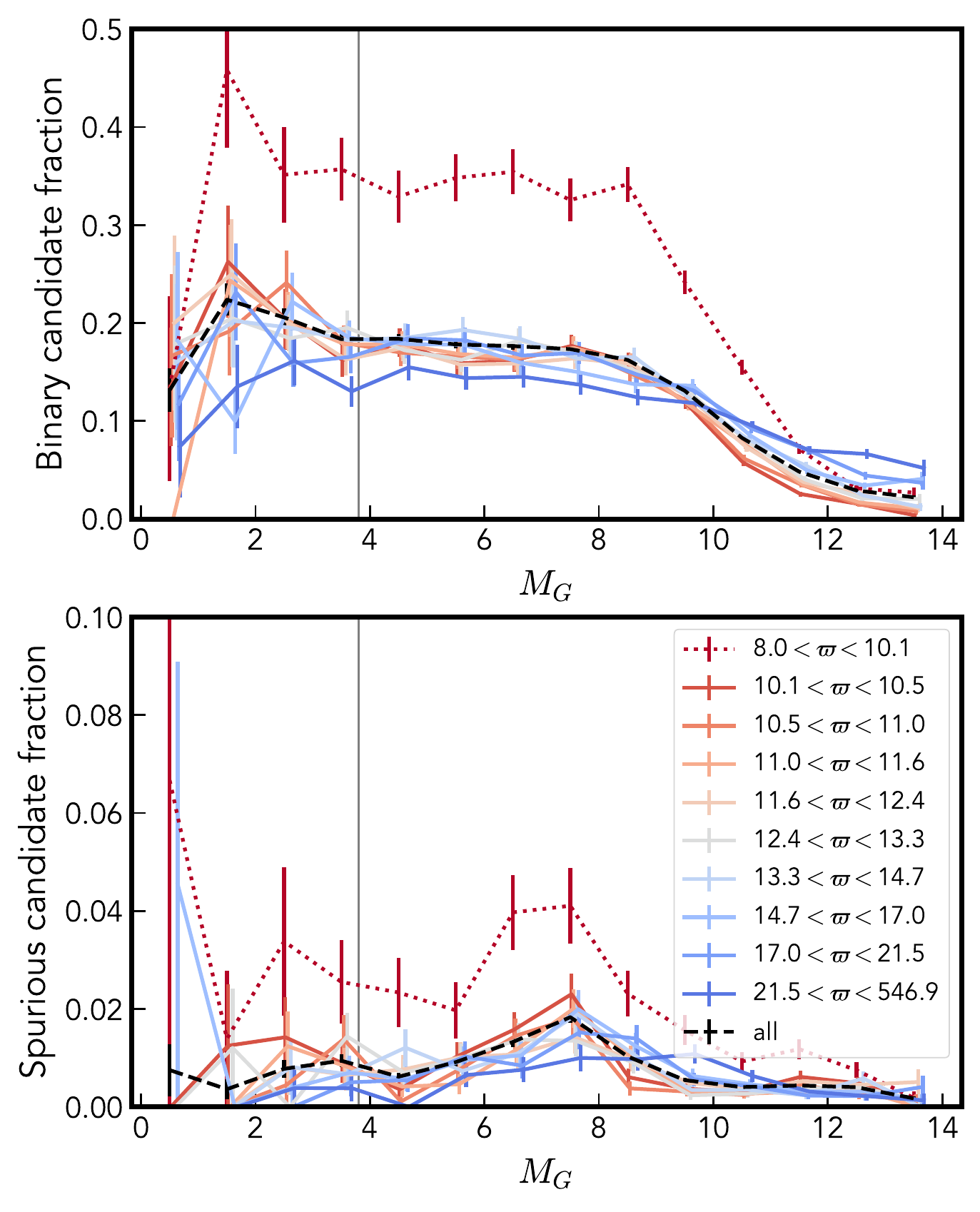}
\caption{The fraction of candidate unresolved binaries and spurious sources along the Main Sequence (including the Young Main Sequence) as a function of distance ranging from nearby sources (blue) to most distant in the dataset (red). The average across the whole population is shown as a black dashed line. Sources are included in GCNS if they are plausibly within 100 pc, allowing some sources with $\varpi<10$ (often with large $\sigma_\varpi$) - which are included in these plots but denoted with a dotted line to show that they're a disparate and perhaps unreliable population. The vertical line shows the transition from our MS to YMS classes.}
\label{binfrac_mag}
\end{figure}

Much like in Figure~\ref{gaia_binfrac}, we can ask how the unresolved binary fraction varies for MS stars, but now as a function of absolute magnitude - a reasonable proxy for stellar mass. This is shown in Figure~\ref{binfrac_mag}, splitting the sources into bins of different distances. Starting with the sample as a whole, we see a binary fraction approaching zero for dim stars, increasing to a little over $20\%$ for bright stars. As discussed in the introduction this fraction is consistent with approximately all of these sources being binaries when accounting for the limited period range astrometry is sensitive to.

The spurious fraction is low ($\lesssim2\%$) throughout, peaking at $M_G \sim 7$. The small number of particularly bright sources introduces significant noise, there may be an uptick for the largest MS stars or it may remain approximately flat.

Looking at the behaviour as a function of distance we see the impact of two selection effects:
\begin{itemize}
\item unresolved binary motion - for less massive dimmer stars, the requirement that we observe a significant fraction of an orbital period to have a significant $LUWE$ limits the maximum semi-major axis of an orbit, and thus the significance of the binary astrometry. Thus we are more sensitive to these low mass binaries when they are closer, as seen at high magnitudes.
\item resolved binaries - if the two stars are sufficiently separated, it becomes possible for Gaia to partially or fully resolve the two sources. This may lead to an extended image and a compromised fit and at larger separations there will be two entries (with low $LUWE$) in the catalogue and thus the binary will not appear in this analysis. This may explain the dearth of particularly nearby binaries, particularly for $4\lesssim M_G \lesssim 8$.
\end{itemize}

The behaviour of partially resolved sources and their astrometric indicators in \textit{Gaia} (particularly \texttt{IPD\_GOF\_HARMONIC\_AMPLITUDE} and \texttt{IPD\_FRAC\_MULTI\_PEAK}) are discussed in more detail in Appendix \ref{IPDbinary}

\begin{figure*}
\centering
\includegraphics[width=0.98\textwidth]{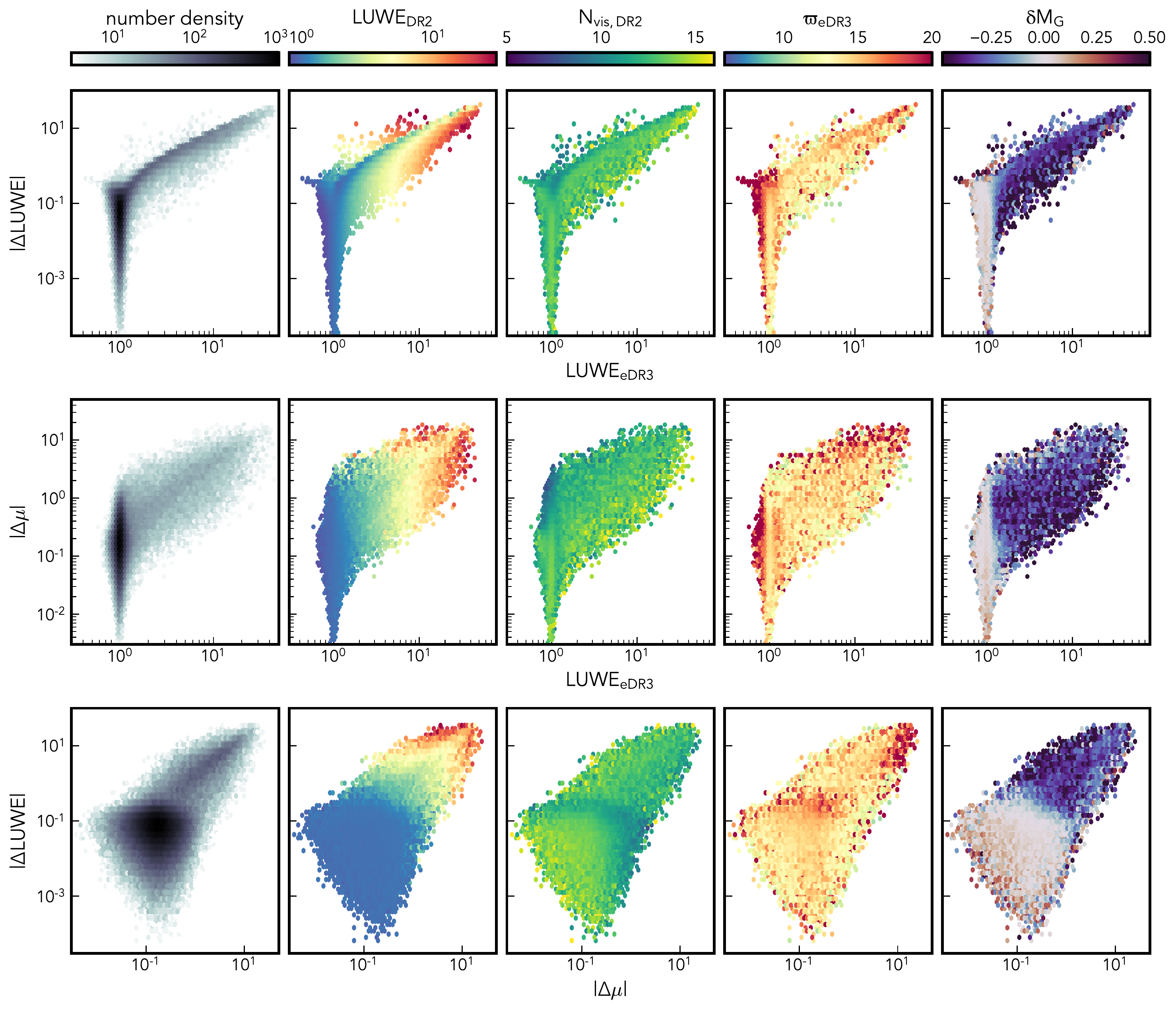}
\caption{Comparing the change in $LUWE$ between Gaia eDR3 and DR2 for MS sources, as a function of $LUWE$ and proper motion anomaly. We show the overall population density (left) and a selection of properties inferred from observation: the DR2 $LUWE$, the number of visibility periods (observations spaced by more than 4 days), the parallax and the shift relative to the median MS magnitude as a function of colour. This figure is the analogue of Figures 12, 13 and 14 of P+21, where we showed simulated binary systems.}
\label{gaia_ruwe_deltaruwe}
\end{figure*}

In Figure \ref{gaia_ruwe_deltaruwe}, we show the real data equivalents of Figures 12, 13 and 14 of P+21. The agreement between the synthetic binary systems presented in paper I and actual observations is excellent. We see the same major features -- for example, the wing in $\Delta LUWE$ vs $LUWE$ corresponding to systems with periods close to the eDR3 observing window. Even more strikingly, we see the distinct sequence, following a very close to linear relationship, of multiple systems in $\Delta LUWE$ vs $|\Delta \bm{\mu}|$ -- with a clear dichotomy between systems which show minimal astrometric deviations and a population with distinct binary characteristics.

The only major deviation is in $|\Delta \bm{\mu}|$ vs $LUWE$ where we again see a wing-like overdensity for high $LUWE$ sources. However, unlike in the simulated data where this wing has a sharp upper envelope the distribution of observed sources is much broader, suggesting significantly more spread in real values of $|\Delta \bm{\mu}|$ than our simulated sample.

We see a particularly clear behaviour with number of visibility periods. The single source population has a clear gradient toward larger astrometric deviations for less well sampled sources. There is still some variation with parallax, suggesting again that the $LUWE$ correction factor may have an unmodelled distance dependence. Systems consistent with being binaries also tend to be brighter (negative $\delta M_G$) as we would expect for a second luminous companion.

That our candidates behave similarly to our synthetic binaries is a promising sign that our simple metrics for identifying binaries are relatively robust. Though the properties of the simulated systems in paper I are chosen very simplistically, the promising agreement here suggests that in future we may be able to fit more complex parameterised distributions by comparing simulated and observed distributions of astrometric deviations.

\subsubsection{Searching for dark companions}

Astrometric measurements have the capacity to detect motion caused by a luminous source orbiting a massive dark companion, such as a neutron star or black hole. We can look for systems which might host dark companions by looking for large astrometric signals ($LUWE$ and $|\delta \bm{\mu}|$) in systems which show no excess brightness (i.e. small $|\delta M_G|$).

\begin{figure}
\centering
\includegraphics[width=0.49\textwidth]{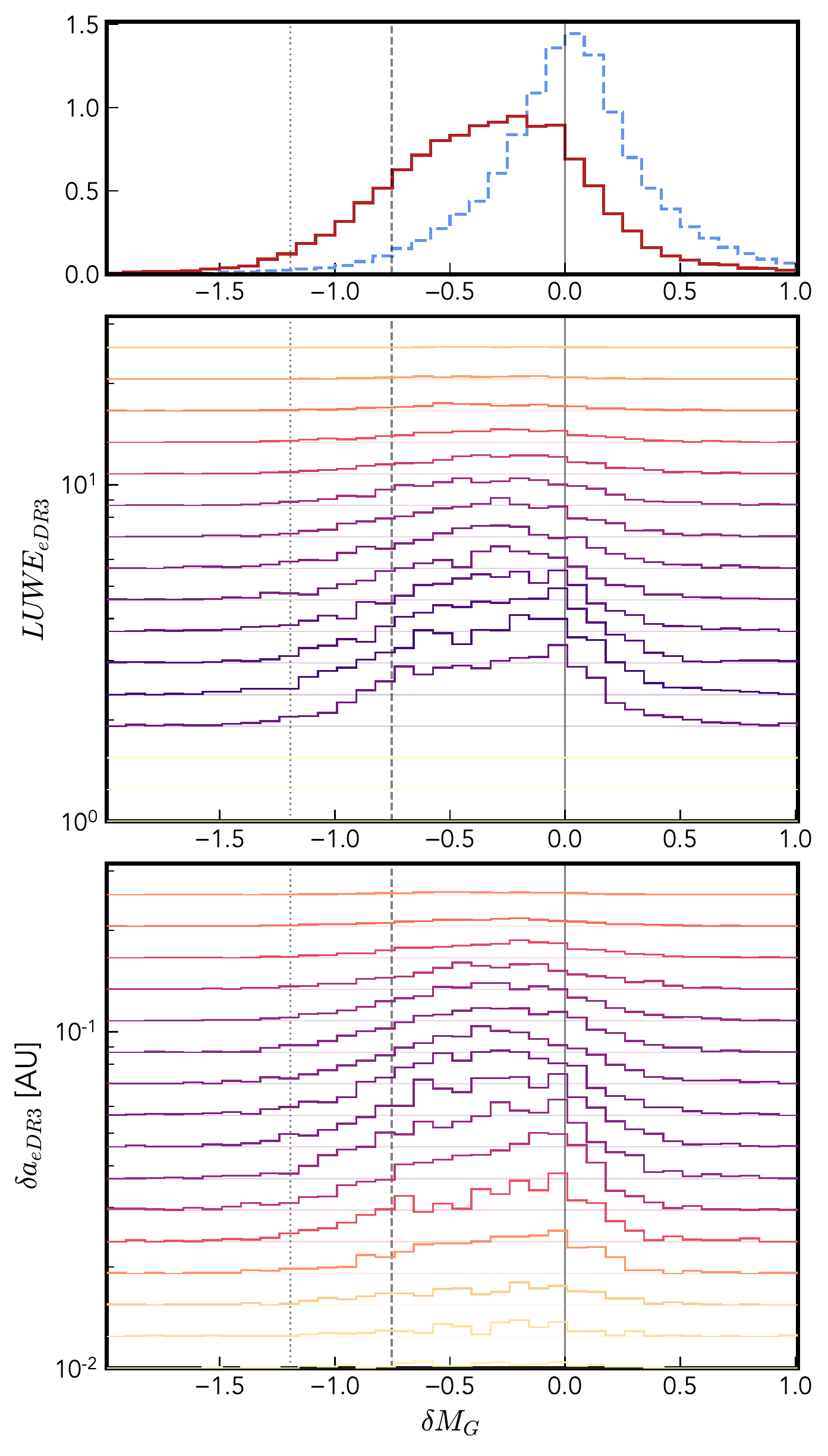}
\caption{Histograms of the magnitude shift ($\delta M_G$, equation \ref{damag}) of MS stars relative to the median (as a function of colour). Top - distribution of all MS binary candidates (red) compared to all other stars which pass our quality cuts (dashed blue). Middle - histograms of candidates in a specific range of $LUWE$, and bottom - of specific $\delta a$, all with the same (linear) y-scaling and colour denoting the number of sources in that $LUWE/\delta a$ bin. Vertical lines show the expected magnitude shift for one, two or three identical stars, as detailed in Figure \ref{msCutout}}
\label{deltamagnitude}
\end{figure}

In Figure \ref{deltamagnitude}, we show the distribution of $\delta M_G$, compared to $LUWE$ and $\delta a$ where
\begin{equation}
\label{deltaa}
\delta a = \frac{\delta \theta}{\varpi}
\end{equation}
is the time averaged projected separation in physical units (au) and
\begin{equation}
\label{deltatheta}
\delta \theta = \sigma_{ast} \sqrt{LUWE^2 - 1}
\end{equation}
is the time averaged angular separation.

We would expect a $\delta M_G$ to be -0.752 for two identically bright stars ($l=1$) and 0 for a completely dark companion ($l=0$). Positive values, which cannot be sensibly converted to a value of $l$, are possible because of our rudimentary method for calculating $\delta M_G$ (comparing to the median magnitude for a given colour) and the inherent uncertainty on both apparent magnitude and parallax. Examining the Figure, we see what appears to be a distribution with a maximum value of $\delta M_G=0$ for single stars, a slight brightness excess for binary candidates and a decline beyond a value of $\sim-0.75$. There may be a very slight change in behaviour around $\delta M_G \sim -1.2$, corresponding to three stars of equal luminosity, but the number of sources is low enough here for noise to begin to dominate.

Whilst we expect $\delta M_G=0$ for dark companions, it could also be a sign of two stars with very different brightnesses (MS stars of very different masses, or a MS+Giant binary) or equally could be contamination by single sources. The fact that the strongest peak at $\delta M_G=0$ is seen in the lowest $LUWE$ systems would fit with either of the latter explanations. At higher $LUWE$, and across all $\Delta LUWE$, the spread of $\delta M_G$ is broad enough that it would be hard to separate a peak at 0 from the rest of the distribution.

True dark companions probably have, as we showed with the simulated systems in P+21, particularly large $\delta a$ values. Looking at the distribution of $\delta M_G$ with $\delta a$, we again see that it's the smaller inferred orbits which have an excess consistent with $l=0$ and for larger orbits the peak is sufficiently broad that picking out an excess at $l=0$ becomes impossible.

The largest takeaway from this exercise is that Gaia eDR3's precision (in $m_G$ and $\varpi$) is not sufficient for these systems to separate sources with different $l$ values, at least not based upon this rudimentary analysis. However, with more careful source selection, data from other telescopes, or a longer observing time (as will be provided in future data releases), identifying sources consistent with a massive dark companion may soon be possible. This is a particularly promising way to start to build a census of the population of low and intermediate mass black holes that we might expect to see in the local Milky Way \citep{Andrews19,Chawla21}. Though the population of large orbit low $l$ binaries cannot be picked out cleanly here, we anticipate being able to resolve these sources soon.

\section{Conclusions}

In this paper, we explore how deviations from single object astrometry allow us to infer the presence of an unresolved binary companion. Specifically, we look at our ability to identify individual binary systems, based on datasets spanning the Gaia DR2 and eDR3 epochs. We use this to pick out candidate binary systems from the Gaia Catalogue of Nearby Stars, a near complete list of astronomical point sources within 100 pc. We find, after some quality cuts, 22,699 candidate binaries, just under 10\% of the sources within this volume.

We use only Gaia data, examining the DR2 and eDR3 results for stars in the GCNS \citep{Smart20}. This sample of just over 300,000 stars within 100 pc has been significantly pruned to remove sources with spurious parallaxes using a random forest. 

The Renormalised Unit Weight Error (RUWE) is the Gaia data product which corresponds to the UWE (the reduced chi-squared of the best fitting 5-parameter single-body astrometric solution). The renormalisation step is performed to account for our limited knowledge about the astrometric error (as a function of colour and magnitude). We show that this renormalisation step, which is applied to the whole Gaia sample, disproportionately re-weights nearby stars (which are dominated in number by distant bright stars of a similar colour). Thus we renormalise the RUWE across just the GCNS sample to give the Local Unit Weight Error (LUWE) with a distribution that peaks at unity. Even after this there is a trend of lower RUWE for nearer sources, suggesting that more careful analysis (especially of larger samples) may require renormalisation to be performed in a series of distance bins.

We remove crowded sources, those for which we cannot identify a corresponding DR2 source, or which do not have a 5 parameter solution in DR2, as well as those with low parallax significance and a high number of bad observations. This leaves around 80\% of the initial GCNS sample.

As in B+20, we show that regions dominated by multiple stars (e.g. above the Main Sequence, MS) stand out clearly when we colour the HR diagram by LUWE, and that most of these show an average increase in LUWE from DR2 to eDR3. There is one region for which the change in LUWE is more often negative, the Sub-Main Sequence (SMS, directly below the MS), which suggests that some of these are not truly multiples but there is another source of astrometric contamination. The regions with high LUWE also show significant proper motion anomaly and parallax shift, consistent with binary systems.

We separate our sources into basic classifications based on colour and absolute magnitude. We then use the very simple classification of $LUWE>2$ and $\Delta LUWE > -LUWE/3$ to select binary candidates in our remaining sample, giving 22,699 sources, whose properties are included in the supplementary data. 

By number, these are dominated by MS sources, but include many examples of each class. These are on average brighter than the median behaviour of the whole sample, with the WD binary candidates in particular seeming to group around 0.75 magnitudes brighter than the median WD behaviour, the increment consistent with two sources of similar luminosity.

We show that Giants and Young Main Sequence (YMS) have the largest fraction of high LUWEs ($\sim 20\%$ above 2), followed by MS stars ($10\%$) whilst LMS stars (the dimmest end of the MS) and White Dwarfs (WDs) have the lowest fraction (1 and 0.5\% respectively). 

It is expected that \textit{Gaia} observations of a population composed of entirely binaries following the period distribution of \citet{Raghavan10} should detect significant astrometric deviations in $\sim 20 - 30 \%$ of systems. Thus our observations are consistent with the majority (or even all) massive stars having companions, agreeing with previous results (e.g. \citealt{Duchene13,Offner20} and references therein).

The very low candidate fraction for dim sources is not necessarily representative of the true multiplicity fraction of these populations. Most Giant binaries will evolve to become WD binaries so we might expect similar true binary fractions. Instead this low candidate fraction is due to the high astrometric error for dim sources. $\sigma_{ast}$ is a strong function of apparent magnitude for sources with $m_G\gtrsim 8$ and thus the contribution of the binary is washed out in any dim population such as WDs and low mass MS stars.

We are able to use this classification to estimate the unresolved binary fraction across the HR diagram. This is lowest close to the middle of the main sequence, highest for particularly bright systems similar to the Sun, and generally high for Giant and YMS sources. Given that binaries can bias parallax measurement, some of the scatter may come from erroneous distances (which then translates to a shift in absolute magnitude). We certainly expect, and appear to see, over-bright photometric binaries sitting above the MS. We also see potential MS+WD binaries which would logically sit at the bluest fringe of the MS. Reversing the condition excluding strongly negative $\Delta LUWE$, we can see where likely spurious high $LUWE$ sources dominate - most notably the SMS.

We also examine the possibility that the high $LUWE$ is caused not by orbital motion of the binary. One such cause is variation in the brightness in one or both sources in an unresolved binary, causing the centre of light to shift over time - often termed Variability-Induced Movers (VIMs, \citealt{Wielen96}). Comparing the relative variation in the flux from each source to the $LUWE$, as a function of position on the HR diagram, we see that many sources are strongly correlated. This suggests binary VIMs may be partially or totally responsible for around 20\% of high $LUWE$ systems. Equally, this may be a sign that these systems are poorly characterised and there may be other sources of their high $LUWE$.

We end by focusing just on the MS, where we can calculate the binary fraction as a function of absolute magnitude (a rough proxy for mass). This tends to zero for the dimmest sources, increases to around $25\%$ above $M_G \sim 7$ and is roughly flat above this. There is some variation with distance, which we can identify with two selection effects. Firstly a dearth of nearby massive binaries (where some fraction of the time we can independently resolve the two sources). Secondly an excess of nearby low-mass binaries (where the small orbit required to have a short enough period for us to see an $UWE$ can only be significantly resolved for close systems).

Finally, we compare the distribution of $LUWE$, $|\Delta LUWE|$ and $|\Delta \bm{\mu}|$ for GCNS MS sources, which shows excellent agreement with the equivalent distributions of simulated sources. Particularly, we see a clear separation between single sources and binary candidates in $|\Delta LUWE|$ and $|\Delta \bm{\mu}|$, and a tendency for sources with high astrometric error to be brighter throughout.

As far as possible we tried to keep our criteria for classification simple and thus we do expect that a significant fraction (though a minority) of our candidates may not prove to be true multiple systems. Radial velocity observations, and in some cases direct imaging or transits, may be able to confirm many of the systems. A detailed comparison between known binary systems and Gaia observations is beyond the scope of this work, though the binary fits provided by the full \textit{Gaia} DR3 results will provide an excellent test of the accuracy of our classification.

\subsection{Beyond GCNS}

By limiting our sample to GCNS (and hence to the local neighbourhood), we have likely excluded the majority of binary candidates in Gaia. There is no reason why a similar analysis cannot be extended out to larger distances in future.

When extending a similar analysis to larger distances the astrometric precision will be reduced by two factors. The first is the linear dependence on parallax of the angular deviation caused by the binary. The second is the increasing astrometric error for dim sources (for $m_G\gtrsim13$). At larger distances we may expect only binaries containing brighter stars to have sufficiently large signals and high enough precision to be detected. 

At larger galactic distances there is also a higher probability of crowding, especially in structures like streams and globular clusters. A more nuanced cut on dense neighbourhoods may be needed, especially as we would expect bright sources in dim fields to be more differentiable than similar brightness sources at the same source density.

\subsection{Beyond eDR3}

The full third \textit{Gaia} data release is planned for June 2022, supplementing the already released astrometric catalog with updated astrophysical parameters, spectroscopic measurements, and many other new or expanded catalogues of data products. Of particular relevance to this work will be a catalog of a few hundred thousand non-single sources, including eclipsing, spectroscopic and astrometric (fitted to the full epoch astrometry) binaries. These will provide an excellent test of the candidates identified in this work as there should be a significant crossover between the two catalogues. This can further inform us about spurious sources and contaminants to our sample. Only the clearest astrometric binaries will be included, and thus we expect only a slim minority of candidates selected by astrometric error to have published fits, meaning that the analysis presented here will remain relevant to the majority of the catalogue.

In Data Release Four the full epoch astrometry will be released. Even here astrometric error will likely be the first indicator of multiplicity, followed up by direct fits for systems with clean and significant binary contributions. In this and future data releases the longer time baseline, larger number of observations, and minor systematic corrections to the pipeline will also increase the number and diversity of detectable systems. As shown in P+21 there is a notable increase in the number of binaries with significant astrometric contributions moving from DR2 to eDR3, especially for longer period systems, and that trend should continue the longer the baseline of the survey.

We hope this work provides a basic framework for such an analysis and is a step to building a much fuller census of binary and higher multiple systems in the Milky Way.

\section*{Data Availability}

The data underlying this article is freely available at \url{https://zenodo.org/record/6053827}.

\section*{Acknowledgements}
We would like to thank the anonymous referee for their helpful and informative questions and comments. We thank Kareem El-Badry and the Cambridge Streams group, including Andrew Everall, Sergey Koposov, Semyeong Oh, Shion Andrews, Jason Sanders and Eugene Vasiliev for their comments, questions and contributions throughout the preparation of this work. ZP would also like to thank Emily Sandford for their help preparing the draft. This work has made use of data from the European Space Agency (ESA) mission Gaia (\url{https://www.cosmos.esa.int/gaia}), processed by the Gaia Data Processing and Analysis Consortium (DPAC, \url{https://www.cosmos.esa.int/web/gaia/dpac/consortium}). Funding for the DPAC has been provided by national institutions, in particular the institutions participating in the Gaia Multilateral Agreement.

\bibliographystyle{mnras}
\bibliography{bib}
\bsp

\begin{landscape}
\begin{table}
\scriptsize\begin{filecontents*}{binarydatashort_a.csv}
sampleid,sourceid,ra,dec,parallax,pmra,pmdec,ruwe,luwe
0,290653322568704,44.44850159775912,0.7917886784873427,17.04575538635254,87.42140197753906,-54.55632019042969,2.638679027557373,2.440086603164673
1,596798591393792,46.437702757259544,1.7117729330493445,10.972148895263672,107.78741455078125,-25.992326736450195,9.984197616577148,9.099285125732422
2,774472798501120,46.75561090023727,2.134491835812317,12.864794731140137,16.60643768310547,37.94053268432617,23.50187873840332,23.658653259277344
3,1155453577587072,43.726188251009376,1.6485541236514145,20.86394500732422,119.21854400634766,7.349956035614014,7.729955673217773,6.064058780670166
4,1381403216912512,44.47816311926047,2.588188047050787,11.37098503112793,-21.507904052734375,18.63311004638672,5.531933784484863,4.756804943084717
5,1425830358704768,42.7529857336048,2.016595544570301,12.229727745056152,-33.51079177856445,-1.5715813636779785,4.7377753257751465,4.2303338050842285
6,1553614225692544,43.46263795346658,2.5109292046684297,12.141989707946777,-44.907135009765625,-32.2764778137207,4.050170421600342,3.8482019901275635
7,1844263252553216,45.7248593179114,3.2987657831540993,12.270655632019043,79.33505249023438,-29.7020263671875,18.333303451538086,15.811066627502441
8,1990257780894208,44.22755027209574,3.2405624170392286,11.416189193725586,-27.446088790893555,-42.65177536010742,4.111851692199707,3.7960712909698486
9,2109382993763712,44.296836691527666,4.098236887535778,10.610448837280273,29.52808380126953,-17.397111892700195,32.59772872924805,29.351415634155273
10,2184179848615424,44.785225663797746,4.00343415516282,11.653755187988281,120.8603286743164,-117.53762817382812,11.729822158813477,11.633773803710938
11,2517233088211712,48.307934519908706,3.82011414311565,16.796144485473633,276.8719482421875,-229.20306396484375,2.8153135776519775,2.7292873859405518
12,3534796739927424,47.33256163654212,4.5591601636505,12.50866413116455,109.14149475097656,-282.46002197265625,3.59065580368042,3.468526601791382
13,3679313799492864,45.58088078818665,4.446214591925152,17.257450103759766,133.433349609375,-125.89234924316406,15.745722770690918,13.52202320098877
14,3682023923676800,45.466169639224205,4.480997772771263,17.72074317932129,92.65335845947266,-26.9328670501709,5.016003608703613,4.7420759201049805
\end{filecontents*}
\csvautotabular{binarydatashort_a.csv}
\\
\scriptsize\begin{filecontents*}{binarydatashort_b.csv}
deltaluwe,deltara,deltadec,deltaparallax,deltapmra,deltapmdec,RFV,sourcetype
1.3039997816085815,1.2133780728618149e-05,-7.238616035465384e-06,-0.12926357984542847,0.5962109565734863,-2.143641710281372,1.3163660764694214,1
1.3835002183914185,1.559772681503091e-05,-3.6954513689124724e-06,0.963158905506134,0.8082868456840515,0.06799463927745819,9.663139343261719,1
21.121204376220703,2.501367589502479e-06,6.95625931257382e-06,-1.2268016338348389,-8.662774085998535,-8.19438648223877,2.0525095462799072,1
4.291528701782227,1.6611085811746307e-05,1.4397426184586948e-06,0.5438024401664734,-0.5632653832435608,-1.924817442893982,3.270106315612793,1
2.2330076694488525,-2.9426423679979052e-06,2.5844888114079367e-06,0.2802560329437256,0.4367223083972931,-0.05177190899848938,1.7056318521499634,1
1.8142242431640625,-4.796161192643922e-06,-1.200864545580771e-07,-0.10230546444654465,-0.23160023987293243,0.16941066086292267,4.447958946228027,1
0.3312137722969055,-6.206040779943578e-06,-4.192419055470964e-06,0.04002068564295769,0.5797040462493896,-1.4314202070236206,3.362936496734619,1
10.787446975708008,1.1371902473911177e-05,-3.971669684688095e-06,2.366105079650879,0.9908090233802795,-1.3263587951660156,8.161224365234375,1
0.43921294808387756,-3.759844730666373e-06,-5.922452146478463e-06,-0.2801806926727295,0.4230450987815857,0.0128471739590168,2.0188968181610107,1
28.1783504486084,4.616316800820641e-06,-4.430884928297019e-06,-0.6348437070846558,7.270049095153809,9.334968566894531,0.684516966342926,1
10.419553756713867,1.6793524991953745e-05,-1.5671794244553894e-05,-0.14315208792686462,-1.488509178161621,-3.8266570568084717,0.7231729030609131,1
1.578137755393982,3.8562528061447665e-05,-3.176662721671164e-05,0.13272123038768768,-0.46624377369880676,-0.2568559944629669,8.854113578796387,1
1.2200771570205688,1.5215218809316866e-05,-3.922378527931869e-05,0.1977192759513855,0.192840576171875,0.04395127668976784,0.9772591590881348,1
11.737039566040039,1.8131204342353158e-05,-1.8813470887835138e-05,-1.4383928775787354,0.010069962590932846,6.123011589050293,1.0733805894851685,1
3.5581629276275635,1.281581717194058e-05,-3.6665921925305156e-06,-0.04745425283908844,-0.9508036971092224,-0.524552583694458,0.9475558400154114,1
\end{filecontents*}
\csvautotabular{binarydatashort_b.csv}
\\
\scriptsize\begin{filecontents*}{binarydatashort_c.csv}
gmag,bpmag,rpmag,absmag,bprp,dmag,dtheta,da
15.254454612731934,16.938892364501953,14.001383781433105,11.412535667419434,2.9375085830688477,0.0019946799147874117,0.6220189929008484,0.036491136997938156
14.928210258483887,16.39621353149414,13.640143394470215,10.129669189453125,2.756070137023926,-0.7001338601112366,2.188009738922119,0.19941487908363342
9.74286937713623,10.144306182861328,9.175890922546387,5.289883613586426,0.9684152603149414,-0.14311598241329193,3.704251289367676,0.28793707489967346
13.044758796691895,14.452713966369629,11.878026008605957,9.641740798950195,2.574687957763672,-0.6191626787185669,0.8342974781990051,0.03998752310872078
12.760104179382324,13.703254699707031,11.777159690856934,8.039094924926758,1.9260950088500977,-0.20511150360107422,0.6872236132621765,0.06043659523129463
13.823619842529297,14.997700691223145,12.702962875366211,9.260704040527344,2.2947378158569336,-0.09043049812316895,0.6647192239761353,0.054352737963199615
15.921760559082031,17.83298110961914,14.548110008239746,11.343210220336914,3.2848711013793945,-1.070055603981018,1.4042710065841675,0.1156541109085083
13.39859676361084,14.510517120361328,12.162251472473145,8.842935562133789,2.3482656478881836,-0.6871097087860107,2.223361015319824,0.18119333684444427
14.802973747253418,16.187902450561523,13.616090774536133,10.09057903289795,2.5718116760253906,-0.1616198718547821,0.8453951478004456,0.07405231148004532
11.621255874633789,12.318705558776855,10.809361457824707,6.749924659729004,1.5093441009521484,-0.37479692697525024,4.916450500488281,0.46335935592651367
9.936371803283691,10.399691581726074,9.3075590133667,5.268701076507568,1.092132568359375,-0.6582595109939575,1.816389560699463,0.15586301684379578
9.589524269104004,10.126452445983887,8.879202842712402,5.715572357177734,1.2472496032714844,-0.6978915929794312,0.39796534180641174,0.02369385026395321
15.798460960388184,17.348630905151367,14.585768699645996,11.284515380859375,2.762862205505371,0.432343065738678,1.192667007446289,0.09534727036952972
13.762677192687988,15.086170196533203,12.624141693115234,9.94756031036377,2.4620285034179688,0.05622877553105354,2.1334028244018555,0.12362214177846909
10.57829761505127,11.189454078674316,9.835286140441895,6.820707321166992,1.3541679382324219,0.11393364518880844,0.6824606657028198,0.0385119691491127
\end{filecontents*}
\csvautotabular{binarydatashort_c.csv}
\caption{\label{binarydata} The first 15 entries in \texttt{binarydata.ecsv} (ordered by sourceid). All parameters refer to their eDR3 value, though DR2 values can be inferred for many quantities when combined with the change in the parameter from DR2 to eDR3: $delta\_$. The $ra$, $dec$ and their variation between data releases are given in degrees. All other astrometric parameters are given in $mas$ or $mas \ yr^{-1}$. The $sourcetype$ denotes the type of star (as chosen by their position on the HR diagram): MS-1, YMS-2, Giant-3, WD-4, LMS-5, SMS-6. The final three categories are $\delta M_G$ (equation \ref{damag}), $\delta \theta$ (equation \ref{deltatheta}) and $\delta a$ (equation \ref{deltaa}).}
\end{table}
\end{landscape}

\appendix

\section{Quality cuts on GCNS data}
\label{qualityCuts}

\begin{figure*}
\centering
\includegraphics[width=0.98\textwidth]{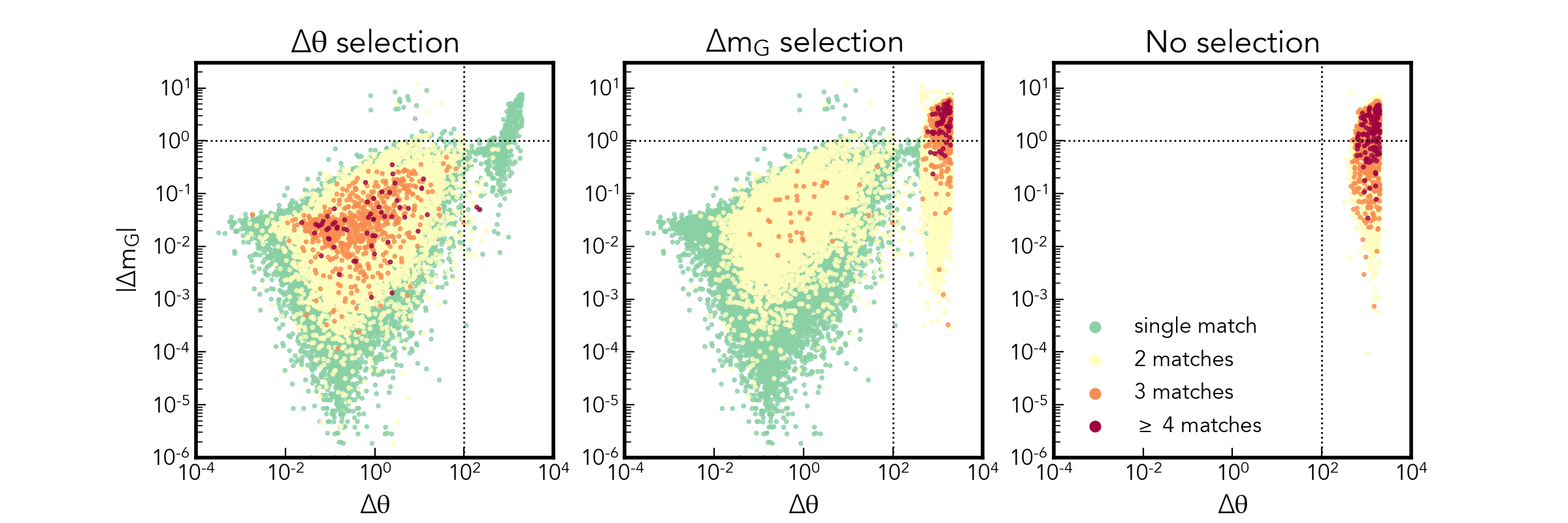}
\caption{The angular distance (in mas) and apparent magnitude difference of sources matched between the Gaia DR2 and eDR3 samples. Points with a single compatible match are shown in green, with higher multiplicities in yellow, orange and red. In the left hand column we select the match based on minimum angular distance, in the middle column we select by minimum magnitude difference, and in the rightmost column we show sources not selected by either matching criterion. The vertical ($\Delta \theta = 100$ mas) and horizontal ($|\Delta m_G|=1$) dotted lines show the proposed limits for a match to be considered valid (see text for more details)}
\label{matchSelection}
\end{figure*}

\begin{figure}
\centering
\includegraphics[width=0.49\textwidth]{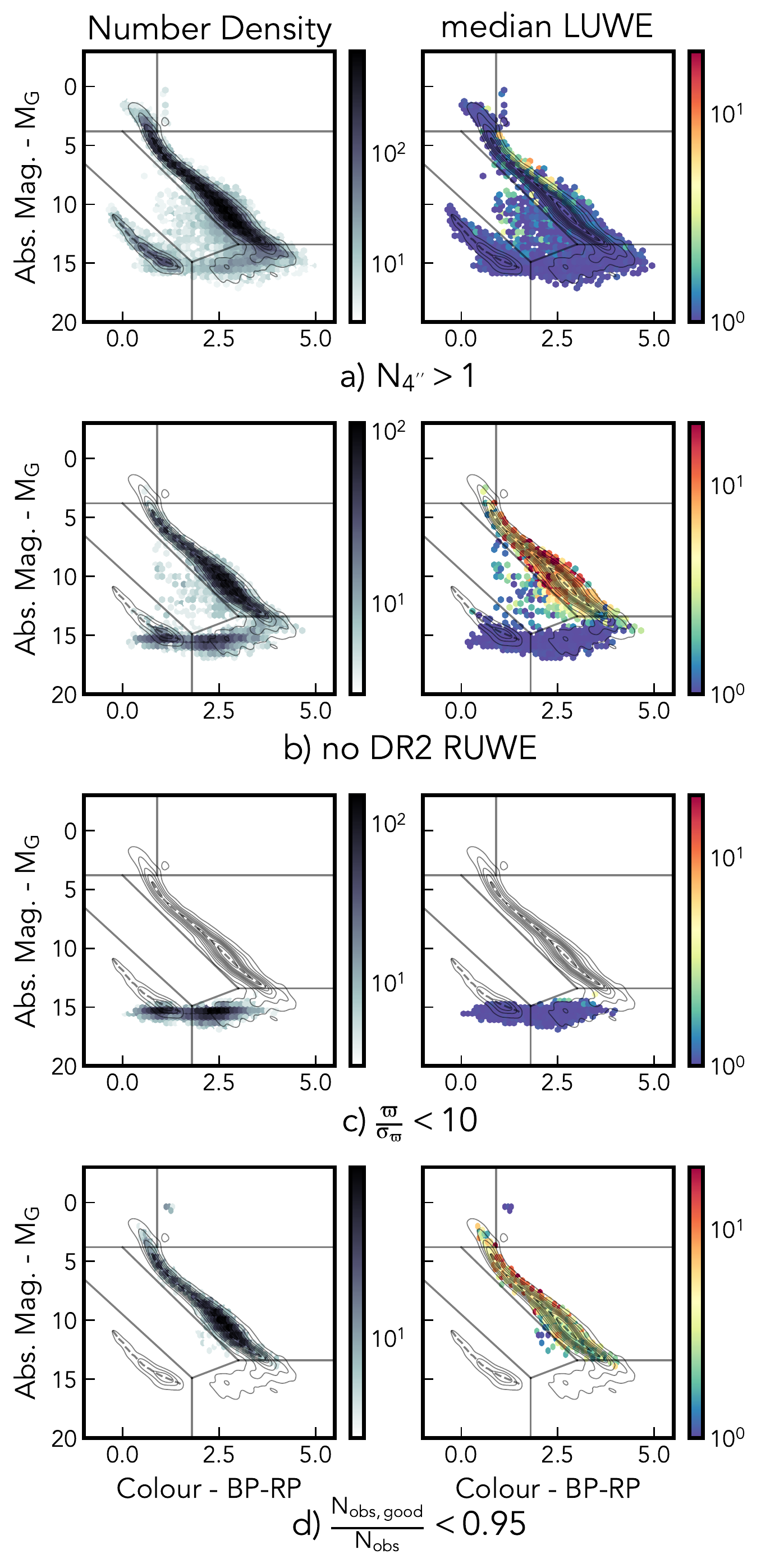}
\caption{The properties of sources removed by our quality cuts shown on the eDR3 HR diagram (see Figure \ref{hrRuwes3} for more details). Only classifications (a) through (d) are shown, as sources failing criterion (e) cannot be plotted in this space and the number of sources cut out by (f) is minimal.}
\label{hrCuts}
\end{figure}

\begin{figure}
\centering
\includegraphics[width=0.49\textwidth]{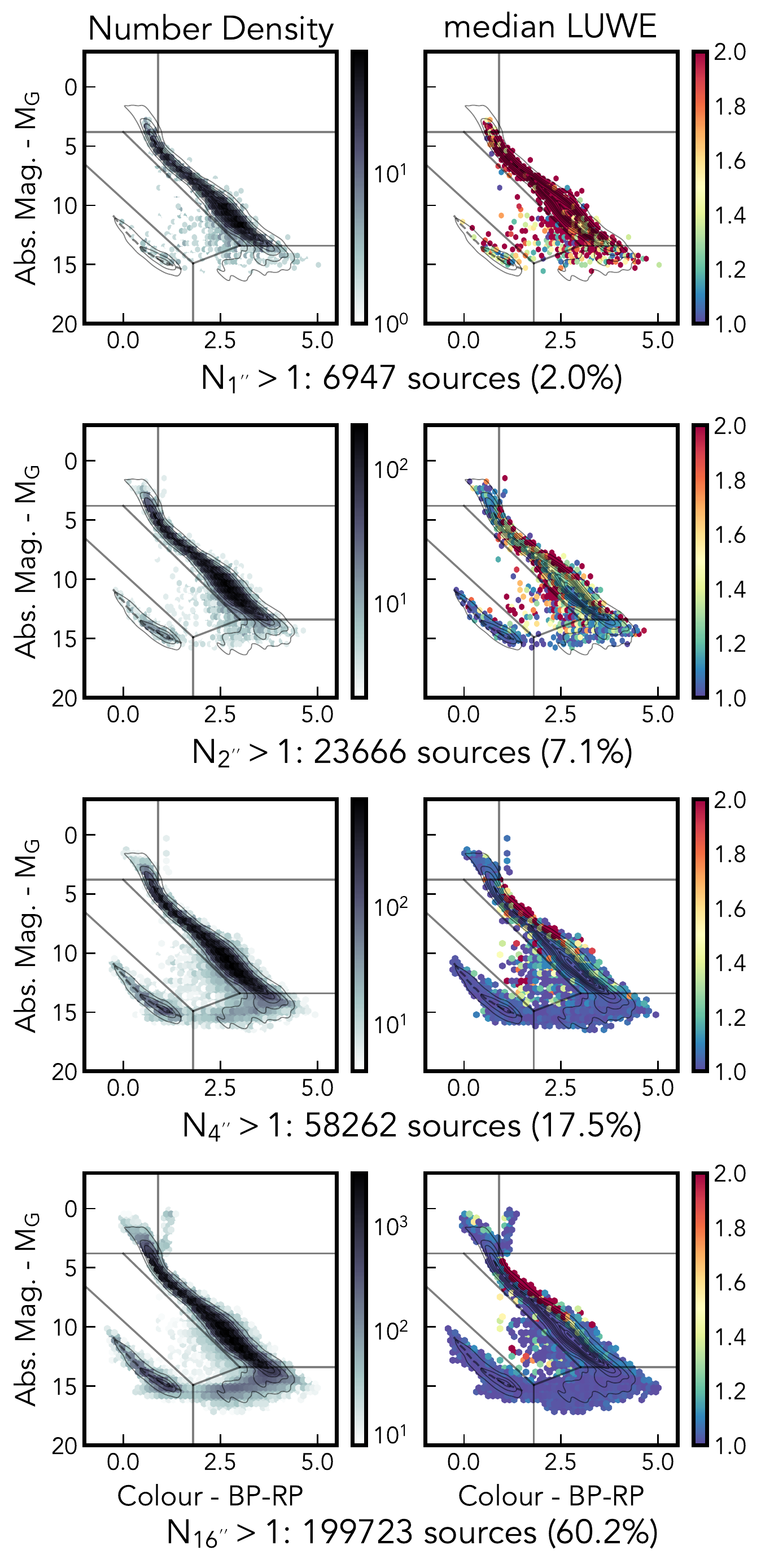}
\caption{Sources with more than one neighbour within $n$ arcseconds ($N_{n"}$) showing the effect of removing sources susceptible to bad astrometric measurements due to blending. Unlike in other plots in this paper bins with at least 1, 2, 4 and 8 sources are shown for neighbours within 1, 2, 4, and 16 arcesond cuts respectively.}
\label{hrCutsCount}
\end{figure}

Before we can investigate the properties of the sample in detail we want to remove sources which are likely to be contaminated or otherwise inaccurately measured. In the interests of understandability of our results we'll apply only a small number of cuts and detail the effects of making each in this section.

The Gaia catalogue of Nearby Stars is constructed such as to remove erroneous samples, using a random forest algorithm classifying using the astrometric and photometric Gaia eDR3 data, and trained on sources with negative parallax. Thus it should contain relatively few bad sources, however there are some potential flags of unreliable sources it doesn't have access to (such as crowding of nearby sources) and others upon which we may want to make more stringent cuts (like the signal to noise ratio of the parallax).

\subsection{Matching between DR2 and eDR3}

As we're interested in the change between the astrometric properties of a source over time one of the first steps we must perform is matching sources between the Gaia DR2 and eDR3 catalogues. The basic tool to perform this matching is provided by the Gaia collaboration in the form of the \texttt{dr2\_neighbourhood} auxiliary table. Here the DR2 data is searched for sources within $\sim$1 arcsecond of the expected position based on eDR3 data - where proper motion data exists for the eDR3 source (a condition we'll apply later) this is used to propagate the motion back to the previous epoch.

However it is possible for multiple sources to reside arbitrarily close to the expected position based on the eDR3 data, and whilst this is generally rare there are many cases where we must choose the best matching candidate.

To do this we turn to the change in apparent magnitude $\Delta m_G =m_{G,eDR3}-m_{G,DR2}$ and angular separation $\Delta \theta$. Both quantities are recorded in the \texttt{dr2\_neighbourhood} table and for a good match we'd expect both to be close to zero.

in Figure \ref{matchSelection} we show the effects of selecting matches based upon the criterion of either minimising angular or magnitude differences. As we might expect an angular selection performs best, with most selections also corresponding to small magnitude differences, thus this is what we use to match our catalogues throughout the rest of this work. In comparison selecting on low $|\Delta m_G|$ selects many sources with large $\Delta \theta$ which are unlikely to be a true match.

A matching entry between the two catalogues is not guaranteed, and we see there is a significant number of sources which don't seem to be valid matches, even for sources with only one possible match. Thus we will also apply a later quality cut of $\Delta \theta < 100$ and $|\Delta m_G|<1$ as indicated by the dotted lines in the figure.

\subsection{Removing sources}

We apply 6 quality cuts to the GCNS sources to try and eliminate most sources of spurious observations. Especially when preparing a sample to search for multiple star systems we do not want to be over-zealous here, as many of the indicators of a poor quality source can be caused by (and serve as evidence for) a binary system. Equally we don't wish to eliminate too large a fraction of the sample. Thus we apply these cuts which eliminate a total of 71,148 sources (21\% of the sample):
\begin{enumerate}[(a)]
\item $N_{4'',eDR3}==1$ - only a single Gaia source within a 4 arcsecond window (58,262 sources removed)
\item $RUWE_{DR2}$ exists - requires a 5 parameter astrometric solution in DR2 (11,170 sources removed)
\item $\frac{\varpi_{eDR3}}{\sigma_{\varpi,eDR3}}>10$ - reasonable signal to noise ratio for the parallax (6,478 sources removed)
\item $\frac{N_{good obs,eDR3}}{N_{obs,eDR3}}>0.95$ - few astrometric observations are excluded from the fit (5,779 sources removed)
\item Valid photometric measurements - we specify that there must be values for $m_G,\ m_{BP}$ and $m_{RP}$ in both DR2 and eDR3 (10,625 and 7,005 sources removed respectively)
\item $\Delta \theta < 100$ and $|\Delta m_G|<1$ - a plausible match between the DR2 and eDR3 catalogues, see previous section for more details (616 sources removed)
\end{enumerate}
Many of these cuts will have significant overlaps and thus the total number of sources removed is less than the sum of the contribution from each.

Figure \ref{hrCuts} shows the properties of the sources removed by each cut, so that we can glean some understanding of which sources we've excluded and the effect that will have on the remaining sample.

The largest cut is on the number of neighbours within 4 arcseconds (calculated from the full eDR3 catalogue) - removing sources from across the HR diagram. Many of these will be valid solutions, but crowded regions allow the possibility of multiple measurements of a second source being wrongly included in the astrometric solution. Though the AGIS pipeline generally downweights these spurious measurements some can still be included and drastically alter the astrometric solution. One region particularly effected is the sub main sequence (SMS, between the stellar and WD main sequences) where stellar main sequence stars with erroneously high parallaxes appear in absolute magnitude to be much dimmer than they truly are. Making this selection also excludes all but the widest resolvable binaries, an interesting sample to study but beyond the scope of this work (see for example \citealt{ElBadry21}).

Sources without valid measurements in DR2 (representing by (b) in the figure but also including (e)) make up the next largest excluded category. Because of the absent data it would be impossible for us to analyse these further, but their distribution is interesting. We can separate them into two main catagories: dim sources, which likely just have insufficient number of detections, and very high LUWE sources, which are mostly brighter and span the main sequence and SMS. Some of the latter may also have few observations, whilst other reside in crowded regions - and some may be true multiples excluded from DR2 because of the complications caused by their extra motion and variation. The remaining two cuts (significance of parallax and few observations excluded) fit neatly into the two types of sources excluded by cut (b), most likely for similar reasons. One small exception is the population of giant stars cut out by (d) - perhaps because of a high level of variability.

The only regions of the HR diagram significantly depleted by these cuts is the LMS, the dimmest sources in the sample stretching the capacity of Gaia to make reliable observations, and the SMS. Whilst we believe there are valid sources in this region we also stipulate that this is the population most affected, and perhaps dominated, by spurious sources. Namely brighter and bluer main sequence stars with erroneously high parallaxes (remembering that more distant sources dominate by number for this local sample) scattered down into this intermediate region.

In Figure \ref{hrCutsCount} we also show which sources are removed by successively more stringent cuts on the number of nearby neighbours. More than half of the sources in GCNS have another Gaia source within 16 arcseconds, and the resulting density and LUWE distribution well mimics the whole population. As shown in B+20 there is a significant excess in astrometric noise for sources with neighbours with 2 arcseconds. We use a cut of 4 arcsecnds here, which may be a little over-zealous, but still excludes a minority of sources.

\section{Image Parameter Determination (IPD) binary indicators}
\label{IPDbinary}

\begin{figure}
\centering
\includegraphics[width=0.49\textwidth]{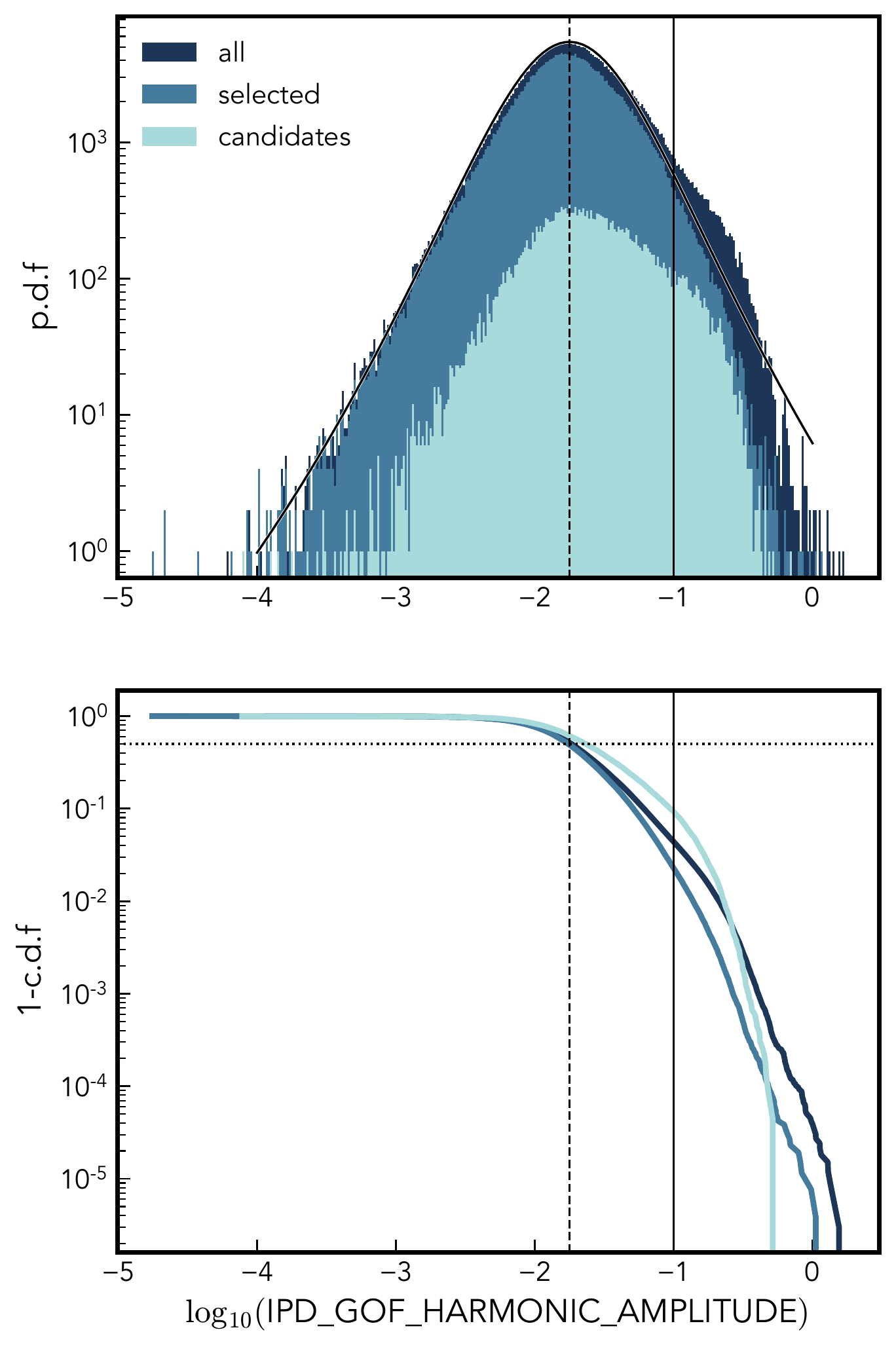}
\caption{The distribution of $\log_{10}($\texttt{IPD\_GOF\_HARMONIC\_AMPLITUDE}$)$ for sources in GCNS. We show the probability distribution function (top) and the survival function (bottom, the fraction of sources above a given value of the parameter) and give the distribution from all sources, those which are remain after our quality cuts, and those which we suggest as candidate astrometric binary systems. Also shown to guide the eye is an approximate fit with a Student's t-distribution with 8 degrees of freedom, a mean of -1.75 (dashed solid line) and a width of 0.33. A significant excess is seen above -1 (vertical solid line) which we attribute to the binary population. We also show a survival fraction of 0.5 (horizontal dotted line) on the bottom plot.}
\label{ipdharmonicDist}
\end{figure}

\begin{figure}
\centering
\includegraphics[width=0.49\textwidth]{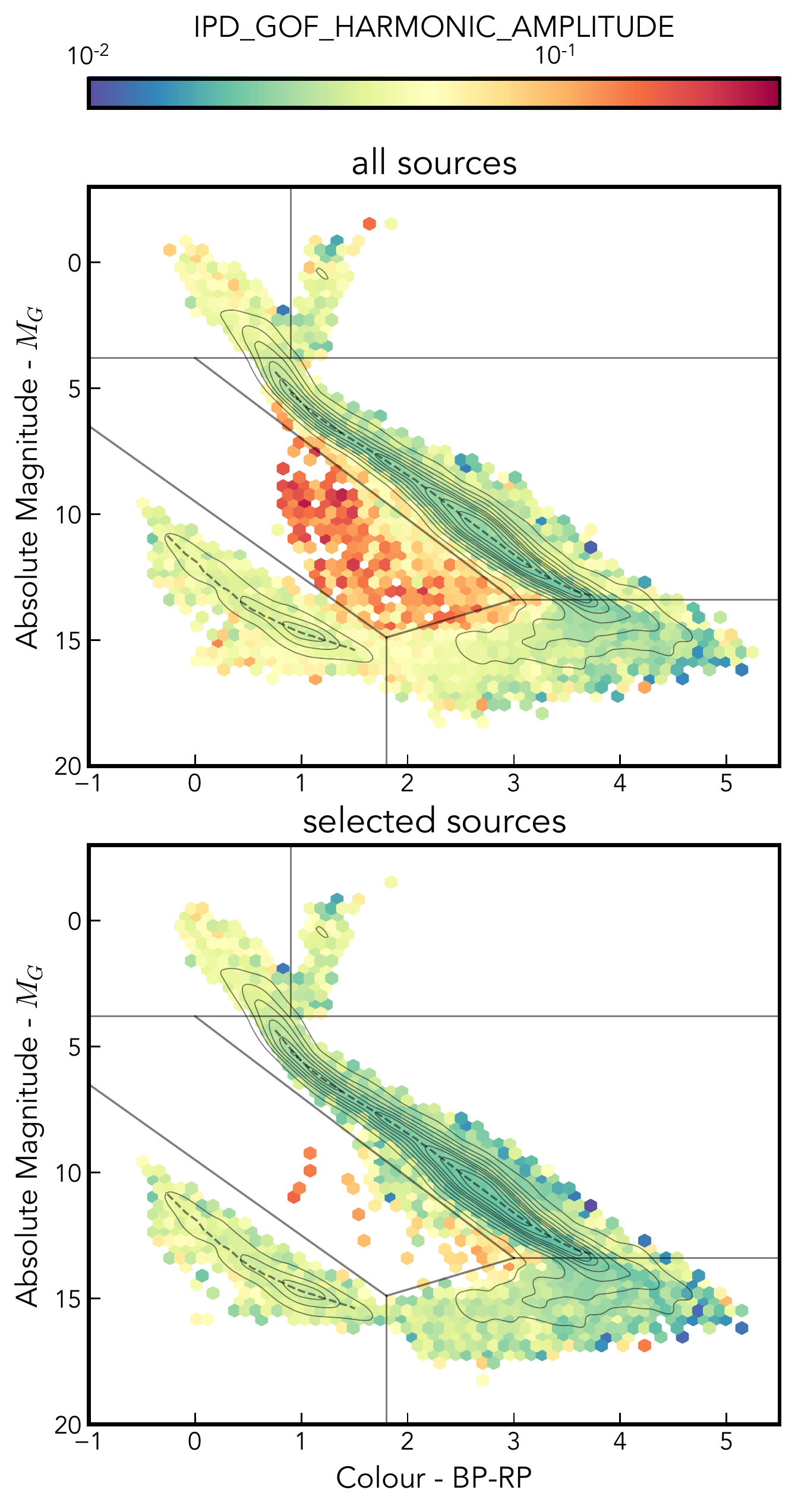}
\caption{The behaviour of \texttt{IPD\_GOF\_HARMONIC\_AMPLITUDE} as a function of position on the HR diagram, shown for all sources and those which pass our quality cuts. Unlike most of the rest of this work we show here the mean (rather than median) value of all sources in each bin.}
\label{ipdharmonic}
\end{figure}

\begin{figure}
\centering
\includegraphics[width=0.49\textwidth]{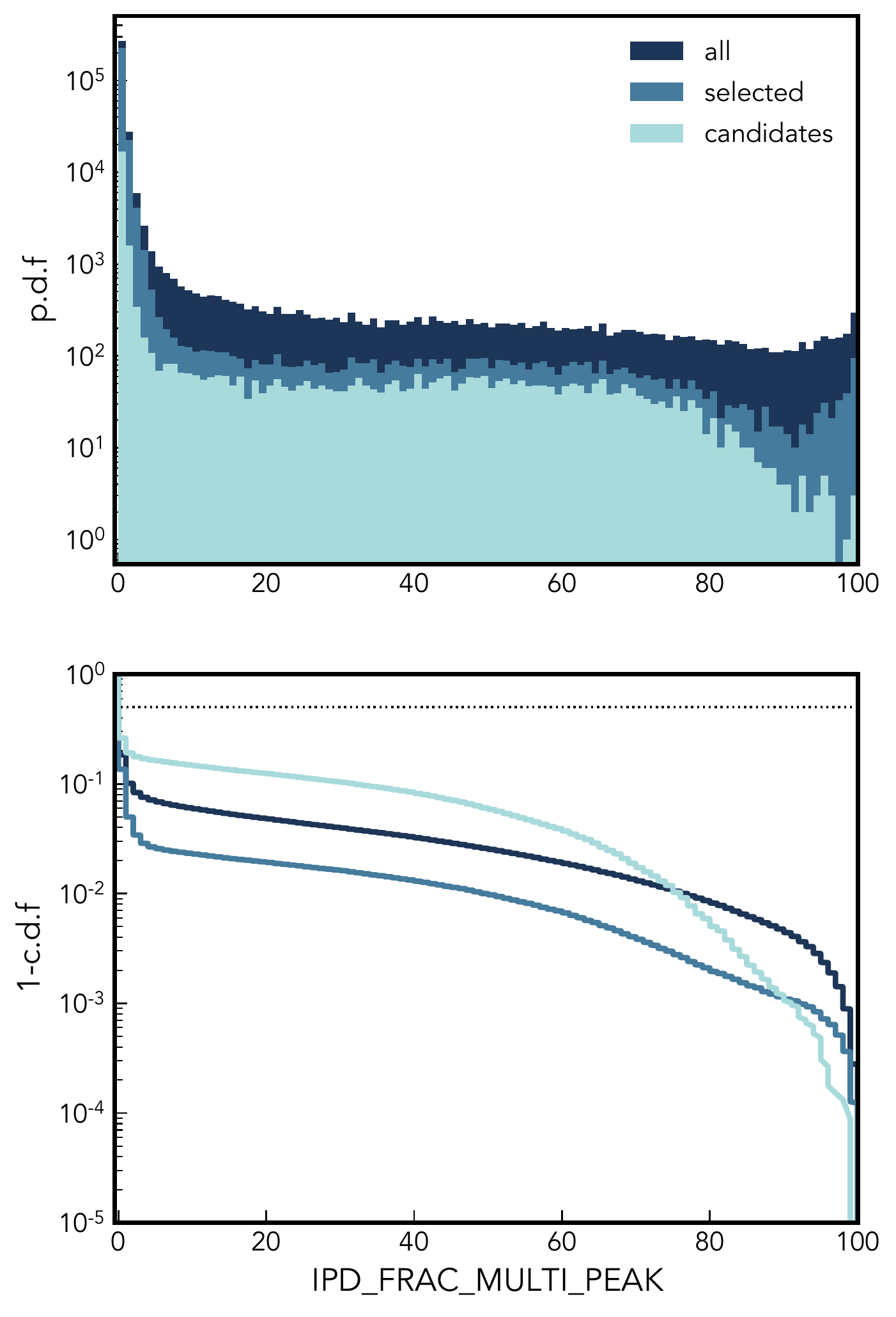}
\caption{Similar to figure \ref{ipdharmonicDist} but now showing \texttt{IPD\_FRAC\_MULTI\_PEAK}. Reported values are integer percentages (between 0 and 100) hence the steplike nature of the c.d.f.}
\label{ipdmultiDist}
\end{figure}

\begin{figure}
\centering
\includegraphics[width=0.49\textwidth]{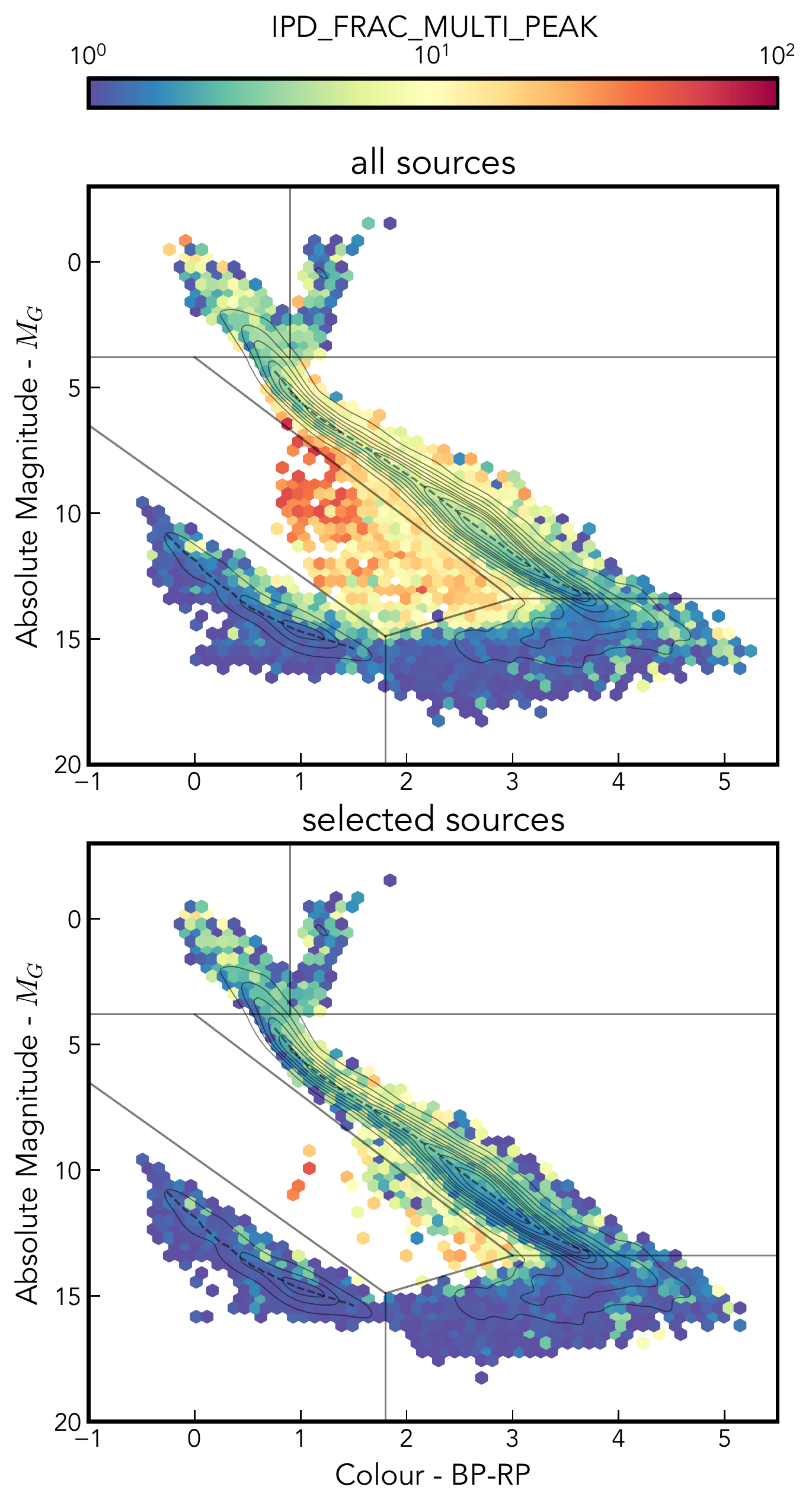}
\caption{The equivalent of figure \ref{ipdharmonic} showing the mean behaviour of \texttt{IPD\_FRAC\_MULTI\_PEAK}}
\label{ipdmulti}
\end{figure}

\begin{figure}
\centering
\includegraphics[width=0.49\textwidth]{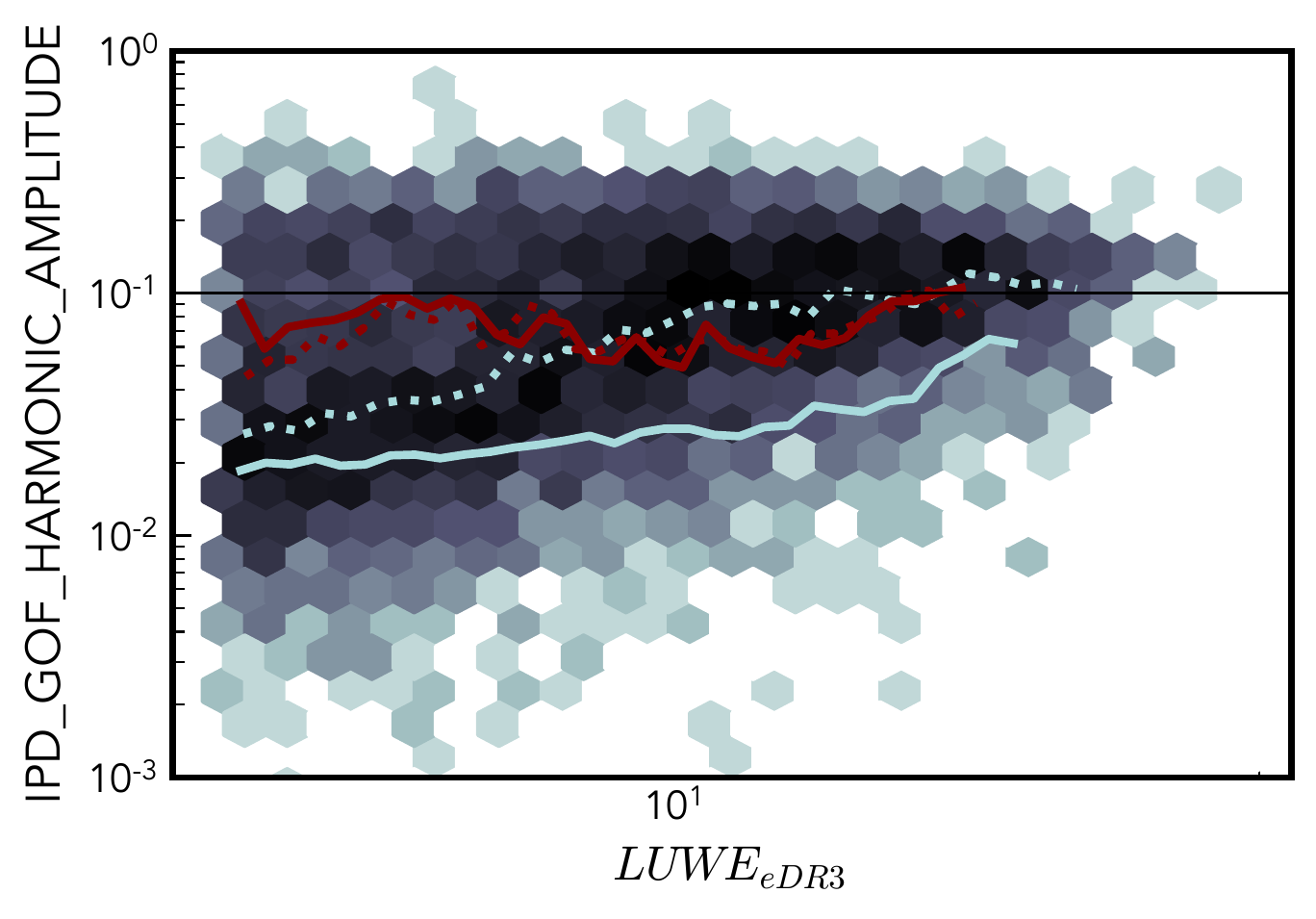}
\caption{The distribution of \texttt{IPD\_GOF\_HARMONIC\_AMPLITUDE} as a function of $LUWE_{eDR3}$ is shown for astrometric binary candidates with \texttt{IPD\_FRAC\_MULTI\_PEAK}$\geq 1$. The median \texttt{IPD\_GOF\_HARMONIC\_AMPLITUDE}, as a function of $LUWE$ is shown as a dotted blue line, and the equivalent without a cut on \texttt{IPD\_FRAC\_MULTI\_PEAK} is shown as a solid blue line. We also show the same relationships for our spurious candidate systems in red.}
\label{ipdharmonicLuwe}
\end{figure}

The Image Parameter Determination (IPD) is an earlier step in the processing of \textit{Gaia's} CCD data. During this phase a point spread function (PSF, or in the case of 1D scans a line spread function, LSF) is used to determine the location of each imaged point-source \citep{Gaia21,Lindegren21}. Much like astrometric error, deviations recorded in this stage may be caused by binarity or other astrophysical contamination, as well as some degree of random noise.

In particular we will focus on two measures, \texttt{IPD\_GOF\_HARMONIC\_AMPLITUDE} and \texttt{IPD\_FRAC\_MULTI\_PEAK}. Both measures are calculated only for scans which pass the \textit{Gaia} quality cuts and for which astrometric data is recorded.

\texttt{IPD\_GOF\_HARMONIC\_AMPLITUDE} records whether the PSF fitting performance has a dependence on scanning angle, which would suggest the source is extended linearly along some particular orientation, as would be the case for a semi-resolved binary. There is an associated angle recorded, \texttt{IPD\_GOF\_HARMONIC\_PHASE}, but here we're only interested in the detecting the presence of a binary and not its properties so we won't concern ourselves with this parameter.

\texttt{IPD\_FRAC\_MULTI\_PEAK} records the fraction of scans in which two distinguishable peaks are visible in the image. Even for a binary where we may expect detectably detached sources there may be some scanning angles (e.g. perpendicular to the binary separation) where only one peak is observed. This measure is sensitive to slightly larger separation pairs than the previous.

As shown in B+20 \textit{Gaia} seems to reliably resolve 2 sources at separations around 2 arcseconds and above (and in this work we use a more conservative cut on sources with neighbours within 4 arcseconds) though they may be distinguishable down to 0.6 arcseconds (\citealt{Lindegren21} Figure 6 and \citealt{Fabricius21} Figures 6 and 7). We would expect these IPD measures to start flagging systems up around one arcsecond separations and down to around 180 mas (the practical lower limit for \textit{Gaia} eDR3, though the optimal \textit{Gaia} LSF width is 110 mas). At larger seperations, if both sources are independently detected with a sufficient number of observations, \textit{Gaia} will record them as separate sources. Thus the IPD measures are sensitive to a population of wider-separation binaries, though we expect some significant overlap, both due to binarity and from the extra astrometric noise introduced by these more extended sources.

They are also relatively sensitive to other extended sources, such as background galaxies, especially the \texttt{IPD\_GOF\_HARMONIC\_AMPLITUDE}. However these should have minimal parallax and are excluded from our sample already. There is also no easy way to tell apart true companions from on-sky neighbours at significantly different distances.

These parameters provide a valuable second vantage point on \textit{Gaia} binaries. We have excluded them from the main body of the analysis, such as to focus on truly unresolved binaries in more detail, but here we will show a similar analysis based on these parameters and their intersection with identifying binaries from astrometric error.

In Figure \ref{ipdharmonicDist} we show \texttt{IPD\_GOF\_HARMONIC\_AMPLITUDE} for different subsets of our data. There is a notable excess for \texttt{IPD\_GOF\_HARMONIC\_AMPLITUDE}$\gtrsim 0.1$ which we attribute to binary sources, although these do not separate from the continuum in the same way as we see for $LUWE$. Our sub-sample of sources which pass our selection criteria (see Appendix \ref{qualityCuts}) show a much smaller excess, but it is clearly present for our binary candidates as well, appearing to verify our stipulation. Around 5\% of all GCNS sources have values greater than 0.1, which drops to 2\% for selected sources and rises again to 10\% for binary candidates. Given that we expect this metric to detect larger separation (and likely higher period) systems the moderate degree of crossover here seems reasonable.

Moving to the HR diagram in Figure \ref{ipdharmonic} we see that certain regions have notably increased \texttt{IPD\_GOF\_HARMONIC\_AMPLITUDE}, most notably the SMS, further confirming that this region may have a significant (perhaps even dominant) population of sources with erroneously high parallax. The excess of sources with high \texttt{IPD\_GOF\_HARMONIC\_AMPLITUDE} is still visible in our selected sample, but now the number is significantly reduced. This is interesting and heartening: we might have expected these sources to be a significant contaminant but our quality cuts (most likely removing sources with nearby neighbours) has been very effective at cutting out these suspected spurious observations. It is not certain whether that is because \textit{Gaia} has two entries for each component of these binaries, or because they are in generally dense regions excluded by our cut on neighbours. It would seem reasonable to make a cut directly on \texttt{IPD\_GOF\_HARMONIC\_AMPLITUDE} but this would exclude some legitimate binaries and the high end of the spread of single star values.

Now we turn to the distribution of \texttt{IPD\_FRAC\_MULTI\_PEAK} in Figure \ref{ipdmultiDist}. The values peak sharply at 0, with around 80\% of all GCNS sources, 85\% of selected sources and 75\% of binary candidates having no scans flagged as containing multiple peaks. Binary candidates generally have higher \texttt{IPD\_FRAC\_MULTI\_PEAK}. This ceases to be true approaching a value of 100, when the two sources are reliably resolved and few observations suffer from an unmodelled blended source, leading to a lower astrometric noise.

The behaviour on the HR diagram, Figure \ref{ipdmulti}, is very similar to that seen for \texttt{IPD\_GOF\_HARMONIC\_AMPLITUDE}. There is now a visible excess above the median MS star and WD sequences as well as among the YMS and (to a lesser extent) Giant stars, consistent with the expectation that these populations are mostly comprised of binary (or higher multiple) systems. 

Finally we can examine the relationship between astrometric error and these measures. In Figure \ref{ipdharmonicLuwe} we show a strong correlation between $LUWE$ and \texttt{IPD\_GOF\_HARMONIC\_AMPLITUDE}, especially for the highest $LUWE$ sources. Selecting the sub-population with \texttt{IPD\_FRAC\_MULTI\_AMPLITUDE}$>$1\%, i.e those which show at least fleeting signs of being two sources, the correlation is even stronger. This seems to confirm that to some extent both of these measures are detecting binarity.

We also show the relationship for sources classified as spurious candidates (with high $LUWE$ but a marked decrease between DR2 and eDR3 incompatible with the behaviour of a binary). These have a much higher \texttt{IPD\_GOF\_HARMONIC\_AMPLITUDE} and no discernible trend, further suggesting that these sources are indeed not binary in nature but are likely some form of blended source or even extended objects with overestimated parallaxes.

\label{lastpage}

\end{document}